\documentclass[11pt]{article}

\usepackage{graphicx}
\usepackage{amsfonts}
\usepackage{amsmath}
\usepackage{amsbsy}
\usepackage{subfigure}
\usepackage{natbib}
\usepackage{color}

\def\vv{{\bf {v}}}
\def\ww{{\bf {w}}}

\def\BB{{\bf {B}}}

\def\EE{{\bf {E}}}

\def\JJ{{\bf {J}}}

\def\RR{{\bf {R}}}

\begin{document}

\title{Three-dimensional magnetic reconnection regimes: A review}

\author{D.~I.~Pontin\footnote{Division of Mathematics, University of Dundee, Dundee, U.K.; ~e-mail: dpontin@maths.dundee.ac.uk}}

\date{}

\maketitle

\begin{abstract}

The magnetic field in many astrophysical plasmas -- such as the Solar corona and Earth's magnetosphere -- has been shown to have a highly complex, three-dimensional structure. Recent advances in theory and computational simulations have shown that reconnection in these fields also has a three-dimensional nature, in contrast to the widely used two-dimensional (or 2.5-dimensional) models.  Here we discuss the underlying theory of three-dimensional magnetic reconnection. We also review a selection of new models that illustrate the current state of the art, as well as highlighting the complexity of energy release processes mediated by reconnection in complicated three-dimensional magnetic fields. 

\end{abstract}

\section{Introduction}
Magnetic reconnection is a fundamental process that is ubiquitous in astrophysical plasmas. It facilitates the release of energy stored in the magnetic field by permitting a change in the magnetic topology in an almost ideal plasma. As such, reconnection is universally accepted to be a key ingredient in the behaviour of many astrophysical plasmas, including the interiors and atmospheres of stars such as the Sun, planetary magnetospheres, accretion disks, and pulsar magnetospheres.

Much of the literature on reconnection  focusses on the two-dimensional problem, due to the theoretical and computational simplifications that this allows.  However, it is now becoming clear that magnetic reconnection in an even weakly three-dimensional (3D) setting is crucially different from the planar 2D case. In this article, we review recent advances in three-dimensional reconnection theory, and the complex picture that is emerging of the possible regimes of reconnection in 3D. This review is by necessity limited and misses a number of important facets of reconnection research. Complementary reviews include those by \cite{priest2000,biskamp2000,zweibel2009,yamada2010}. In Section \ref{topsec} we introduce some key measures and features of magnetic field structure in 3D that are crucial to understanding where and how reconnection operates in 3D, while in Section \ref{recpropsec} we discuss some fundamental differences between 2D and 3D reconnection. In Sections \ref{nonnullsec}-\ref{sepsec} we review the current picture of the various different 3D reconnection regimes, and in Section \ref{obssec} we touch briefly on recent results from large-scale numerical simulations and observations. We finish with a summary in Section \ref{sumsec}.

\section{Magnetic topological and geometrical structures}\label{topsec}
Recent observations and analysis are revealing the complex structure of the magnetic field in astrophysical plasmas. These studies have naturally focussed on solar system plasmas -- in particular the solar atmosphere and Earth's magnetosphere -- because these are environments that we can observe with relatively high spatial resolution. However, it is very likely that this rich structure is present also in other astrophysical bodies.

One crucial step in understanding the behaviour of astrophysical plasmas is to determine just where magnetic reconnection may occur, and therefore what are the likely locations of energy release. Whether in a collisional or collisionless plasma, magnetic reconnection requires the presence of a current sheet. So in order to determine the locations where reconnection can facilitate the release of energy in the plasma, we must understand where current sheets form. (Note: in this article we use the term `current sheet' to refer to any intense, localised current layer -- rather than, as is sometimes the case, reserving the term for singular current structures.)
 In two dimensions, it is well established that reconnection occurs at magnetic X-points, which are prone to collapse to form current layers. However, with the loss of two dimensional symmetry, the number of proposed sites of current sheet formation and reconnection is greatly increased. These proposed sites can be broadly divided into two classes, being either {\it topological} or {\it geometrical} features of the magnetic field. A topological feature is preserved by an arbitrary smooth deformation of the magnetic field, while we will refer to any property that is not preserved by all such deformations  as a geometrical feature.  

\begin{figure}
\begin{center}
(a)\includegraphics[height=5cm]{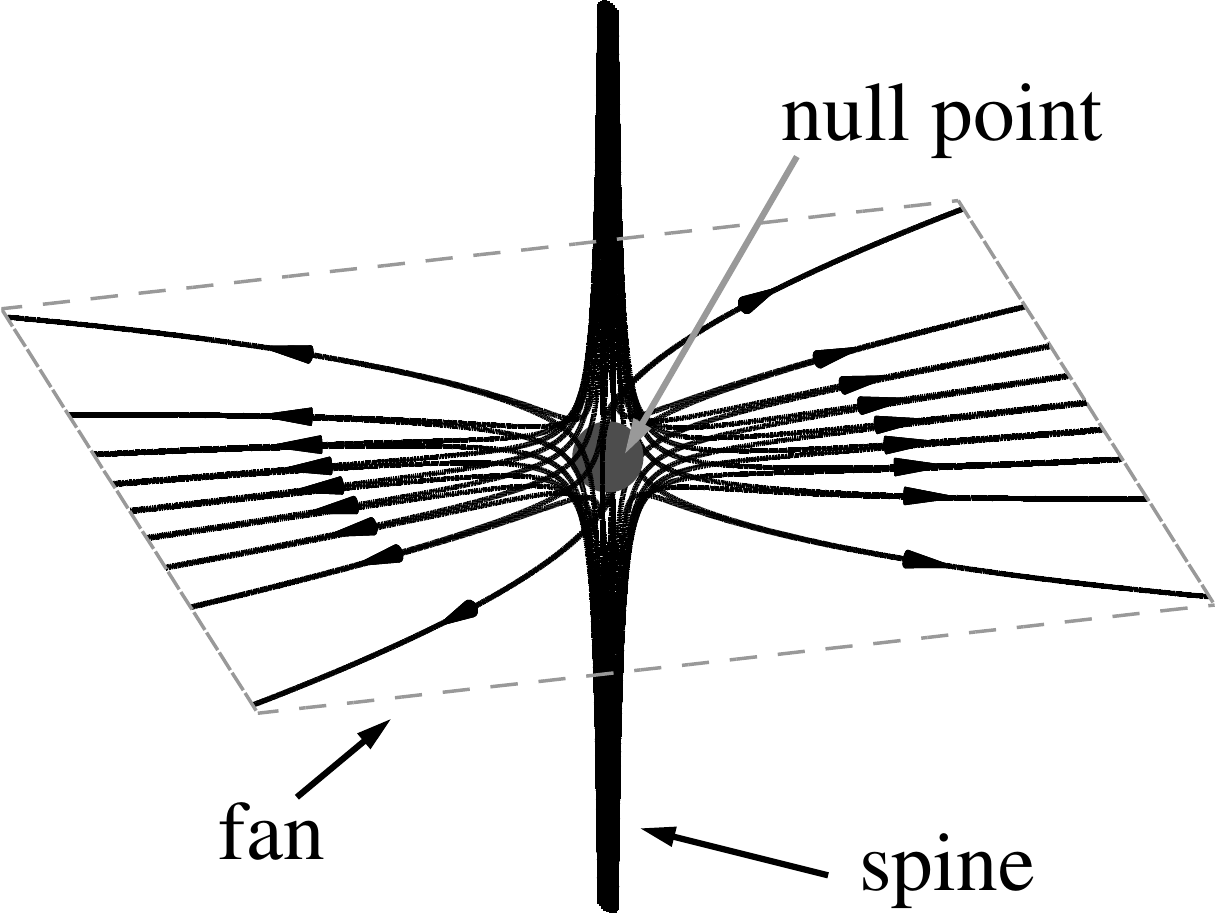}
(b)\includegraphics[height=7cm]{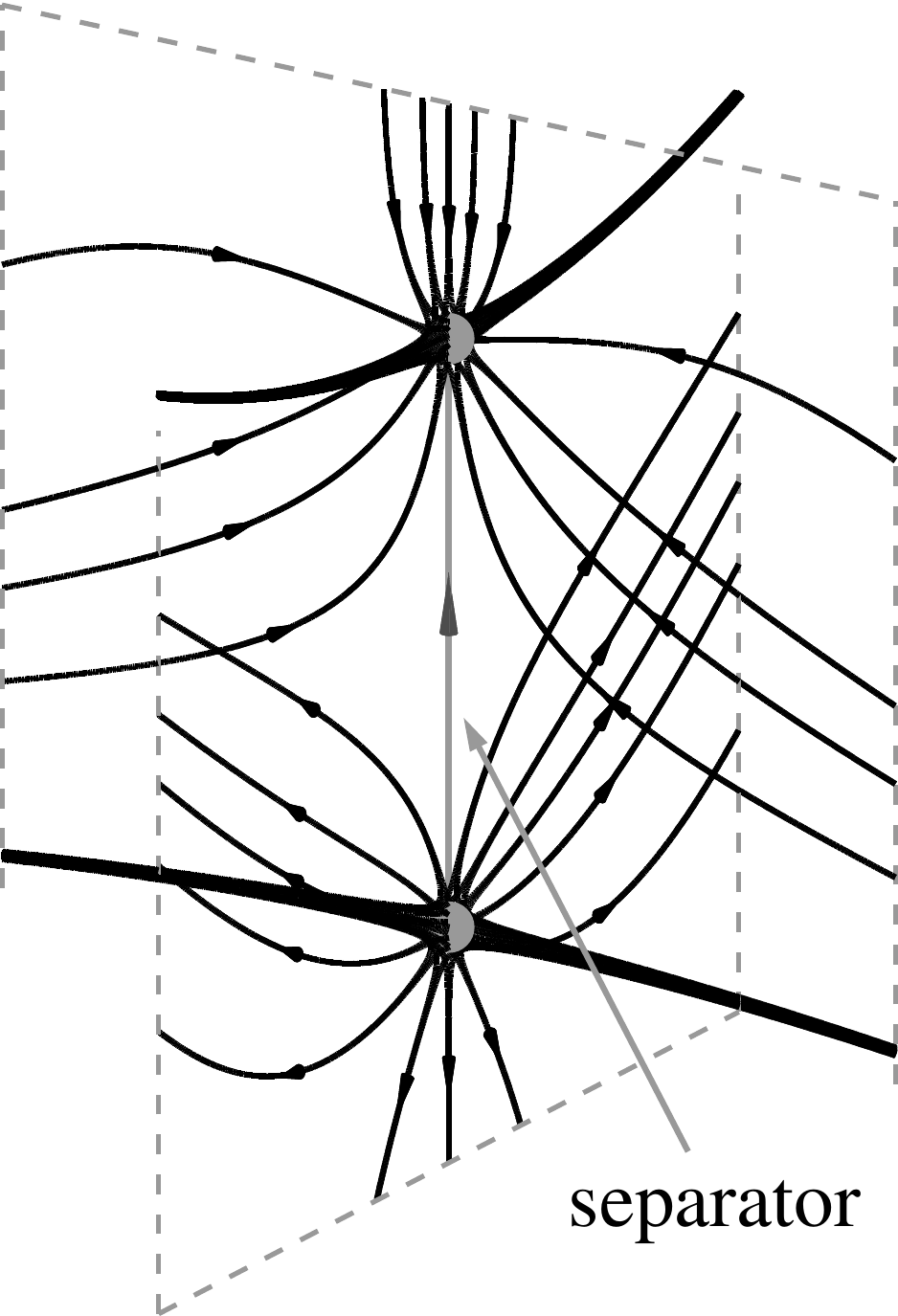}
\end{center}
\caption{Potential magnetic field line structure in the vicinity of (a) an isolated 3D null point, and (b) a generic fan-fan separator}
\label{nullandsep}
\end{figure}

One natural extension from the 2D X-point collapse picture is the idea of current sheet formation at 3D null points -- points in space at which the magnetic field strength falls to zero. Null points are topological features of the magnetic field, and their structure has been studied by a number of authors \citep{fukao1975,lau1990,parnell1996} -- a typical configuration is shown in Figure \ref{nullandsep}(a). Since $\nabla\cdot {\bf B}=0$, magnetic null points must be of hyperbolic type. Their structure is characterised by a pair of field lines that asymptotically approach (or recede from) the null from opposite directions, forming the {\it spine} (or $\gamma$-line) of the null, while field lines recede from (or approach) the null in a surface known as the fan (or $\Sigma$-) plane. This surface is a {\it separatrix} surface -- it separates topologically distinct volumes of magnetic flux.

Another proposed site of current sheet formation in 3D is a separator line -- a field line that runs from one null point to another. 
 In dynamical systems theory this would be called a heteroclinic orbit. Such a field line is defined by the transverse intersection of the fan planes of the two nulls (Figure \ref{nullandsep}(b)) and is therefore  topologically stable.
Nulls and separators were first suggested as possible sites of current sheet formation due to the discontinuous jump in field line connectivity at their associated separatrix structures. Therefore, if one considers the implications of an ideal flow across these separatrix structures in a kinematic model, singularities in the electric field result \citep{lau1990,priest1996}. It is worth noting that these considerations apply equally to closed magnetic field lines, as discussed by \cite{lau1990}. These closed field lines play a crucial role in laboratory plasmas -- however, since the magnetic field in astrophysical plasmas is usually considered to be anchored at the surface of the neighbouring star or planet, closed field lines do not typically take centre stage here. 

A question that naturally arises is: do such null points and separators exist in space plasmas? New observations and theoretical studies suggest that they are abundant. Recent analysis of {\it in-situ} observations by the Cluster mission show the presence of both single nulls and collections of nulls located in the current sheet of the Earth's magnetotail \cite[e.g.][]{xiao2006,deng2009}. Furthermore, the standard model of the magnetosphere contains two null points in the cusp regions joined by separator lines, while clusters of nulls (expected to be joined by separators) have been found in the global magnetosphere simulations of \cite{dorelli2007}.
Turning our attention to the solar atmosphere, the lack of magnetic field measurements in the corona renders direct detection impossible at present. However, increasingly detailed magnetograms at the level of the photosphere permit  extrapolation of the field into the corona. Such extrapolations in quiet sun regions show an abundance of 3D nulls to be present, with a high density at chromospheric levels, falling off exponentially with height \citep[e.g.][]{regnier2008,longcope2009}. Nulls have also been inferred to be present in many flaring and erupting active regions, as discussed later.

Nulls and separators are, however, not the end of the story. In 3D, current sheet formation and magnetic reconnection may also occur in the absence of nulls. 
However, one still requires some mechanism to generate intense current layers in the plasma. In the case of the Earth's magnetosphere, the system is being continually driven by the incident solar wind. This leads to large-scale current structures both at the dayside magnetopause and in the magnetotail (though what triggers these current layers to thin resulting in the onset of fast reconnection is still an open question). In the solar atmosphere, however, the driving of the system is less direct, occurring at the photospheric footpoints, and the question of how this photospheric driving maps eventually to the creation of coronal current sheets is not straightforward to answer. There are certain magnetic field structures that appear to encourage the formation of current layers. One such topological structure is a region of braided magnetic field -- in which field lines are non-trivially linked with one another. A recent series of papers has demonstrated that the small scales associated with the field line mapping in braided fields may lead to a loss of equilibrium, leading to the formation of multiple small-scale current layers \citep{wilmotsmith2009a,wilmotsmith2009b,wilmotsmith2010,pontin2010}. However, current layers may also form in topologically simple fields, for example as the result of some ideal instability. \cite{browning2008} and \cite{hood2009} have followed the evolution of a kink-unstable flux tube in resistive MHD simulations and found a complex array of current layers to form during the subsequent relaxation. 

The structure of a magnetic field may in general be characterised by the mapping generated by the connectivity of magnetic field lines within the domain. In the absence of null points this mapping is continuous. However, it has been proposed that if sufficiently strong gradients are present in this mapping then intense current layers will in general form when the field is perturbed by plasma motions. \citep[e.g.][]{longcope1994,priest1995}. These gradients in the connectivity are characterised by the so-called {\it squashing factor}, $Q$ \citep{titov2002,titov2007}, and regions with high values of $Q$ are usually termed {\it quasi-separatrix layers} (or QSLs). The name stems from the fact that a true separatrix surface may be considered as the limit obtained when a QSL approaches zero thickness and  infinite $Q$. For a detailed review of these ideas, see \cite{demoulin2006}.

A further proposition, originally put forward to explain the heating of the solar corona by \cite{parker1972}, is that any generic footpoint motion will naturally lead to current sheet formation in the corona. This has been investigated by a number of authors using various numerical simulations, without any clear consensus being reached about the nature of the current structures formed (in particular whether they are singular in the ideal limit) or whether they could account for the heating of the solar corona at realistic coronal plasma parameters \citep[e.g.][]{vanballegooijen1985,longcope1994,hendrix1996,galsgaard1996,ng1998,rappazzo2008}.

Here we have given a short (and certainly not exhaustive) introduction describing some possible sites and mechanisms of current sheet formation in 3D magnetic fields. The main focus of this article, however, is not on {\it where} magnetic reconnection may take place in 3D, but rather on the properties of the reconnection process when it does take place.

\section{Fundamental properties of 3D reconnection} \label{recpropsec}

Under what conditions does magnetic reconnection occur in 3D? To answer this question we first require a definition of reconnection in 3D. The most general approach, the one that we follow here, was put forward by \cite{schindler1988,hesse1988} in the framework of {\it general magnetic reconnection}. 
Within this framework magnetic reconnection is defined by a breakdown of magnetic field line and flux conservation, or in other words a breakdown in the magnetic connection between plasma elements. This was shown to occur in three dimensions in general when a component of the electric field parallel to the magnetic field ($E_\|$) is spatially localised in all three dimensions. The change of connectivity, or reconnection rate, is quantified by the maximal value (over all field lines) of
\begin{equation}\label{recratedef}
\Phi=\int E_\| ds
\end{equation}
where the integral is performed along magnetic field lines from one side of the diffusion region (region within which $E_\| \neq 0$) to the other.

It is now being appreciated that the fundamental properties of 3D reconnection are crucially different from the simplified 2D picture. 
These new properties can be understood by considering the implications of the following equation
\begin{equation}\label{idealev}
\frac{\partial {\bf B}}{\partial t}- \nabla\times (\ww \times \BB)={\bf 0},
\end{equation}
which describes the ideal evolution of a magnetic field, where ${\bf w}$ is a flux-conserving velocity or flux transport velocity (which in ideal MHD is simply the fluid velocity ${\bf v}$). If for a given magnetic field evolution a smooth flow ${\bf w}$ exists then the magnetic flux is {\it frozen into} the flow $\ww$, and the topology of the magnetic field is preserved -- this being guaranteed by the condition that ${\bf w}$ be smooth and continuous. (Note that strictly speaking, the topology is still preserved if the right-hand side of Equation (\ref{idealev}) is a non-zero term parallel to $\BB$, say $\lambda\BB$ where $\lambda$ is some scalar field -- see \cite{hornig1996}.)  Using Faraday's law and `uncurling' Equation (\ref{idealev}),
$$
\EE+\ww\times\BB=\RR, \qquad \RR=\nabla\Psi,
$$
where $\Psi$ is a free function. In 2D, the conditions on the existence of $\ww$ are straightforward. 
Assuming that the magnetic and plasma flow fields ($\BB$ and $\vv$) are 2D, then we must have 
$\EE\cdot\BB=0$ and $\RR\cdot\BB=0$, so we can write $\RR=\delta\ww\times\BB$, say, which leads to
$$
\EE+(\ww-\delta\ww)\times\BB={\bf 0}.
$$
So in 2D a flux transporting flow exists everywhere, and is given by 
$$
\ww=\delta\ww+\frac{\EE\times\BB}{B^2}
$$
(see \cite{hornig2001} for a more detailed exposition). It is clear that this velocity is smooth everywhere except at null points (at which $\BB={\bf 0}$) where it is singular unless ${\bf E}={\bf 0}$ there. The above derivation of an explicit expression for $\ww$ relies crucially on the condition that $\EE\cdot\BB=0$. However, in 3D reconnection by definition $\EE\cdot\BB\neq 0$ (see above), and the conditions under which magnetic topology conservation, field line conservation, and magnetic flux conservation hold are much more subtle -- the reader is referred to the papers by \cite{schindler1988,hornig1996,hornig2001,hornig2007a,hornig2007b}.

\begin{figure}
\begin{center}
\includegraphics[width=4.2cm]{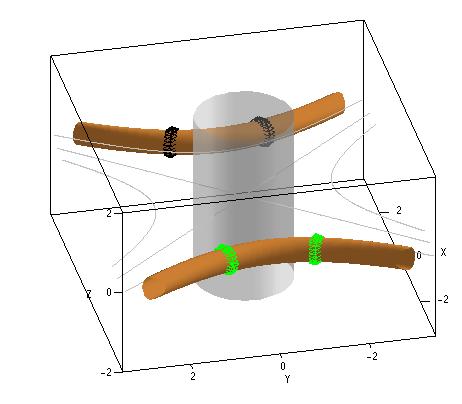}
\includegraphics[width=4.2cm]{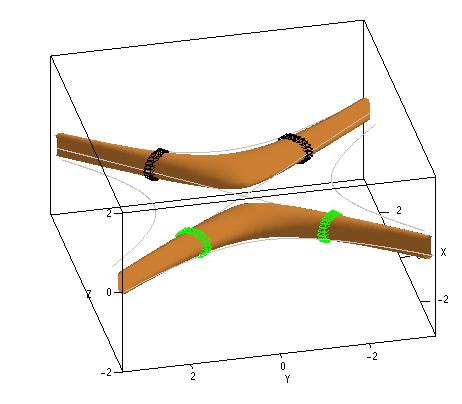}
\includegraphics[width=4.2cm]{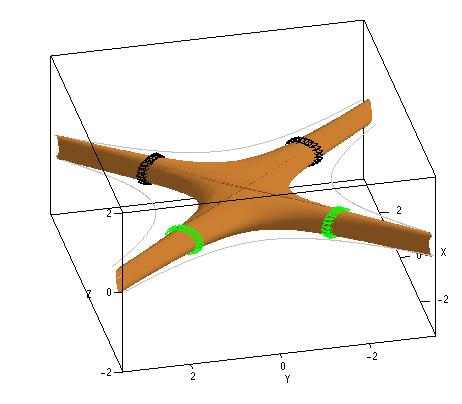}
\includegraphics[width=4.2cm]{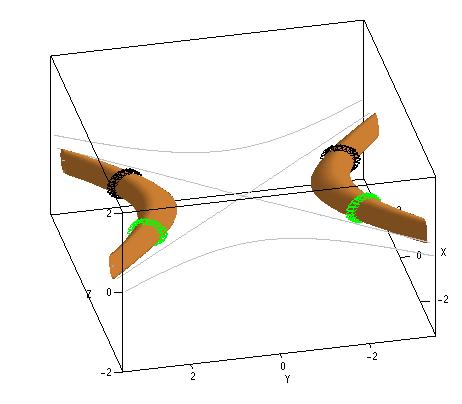}
\end{center}
\caption{Reconnection of two representative flux tubes in the magnetic field $\BB=(y, k^2 x,0)$, with $k=1.2$. A localised diffusion region (shaded surface) is present around the X-point.}
\label{tube2d}
\end{figure}

As demonstrated above, in 2D a smooth flux transporting velocity exists everywhere with the possible exception of magnetic nulls. The magnetic null may be of O-type or X-type: for an O-point with non-zero electric field there is annihilation (or creation) of magnetic flux at the null, while for an X-point there is reconnection of magnetic flux.
The singularity of $\ww$ at magnetic X-points in 2D is a signature of the fact that the reconnection process involves magnetic field lines being cut and rejoined at the X-point. In other words, the field line connectivity changes in a discontinuous manner at the null. Since $\ww$ is smooth and continuous everywhere except at the X-point, field lines evolve as if they are reconnected at this point {\it only}. Hence, the reconnection of magnetic field lines occurs in a one-to-one pairwise fashion at a single point. This 2D reconnection scenario is demonstrated in Figure \ref{tube2d}, where some representative flux tubes are plotted (from ideal comoving footpoints marked grey (green online) and black). The images in the figure show snapshots from an animation of flux tubes reconnecting in a kinematic steady-state solution with $\BB=(y,k^2x,0)$ and a localised diffusion region around the X-point.

\begin{figure}
\begin{center}
\includegraphics[width=5.2cm]{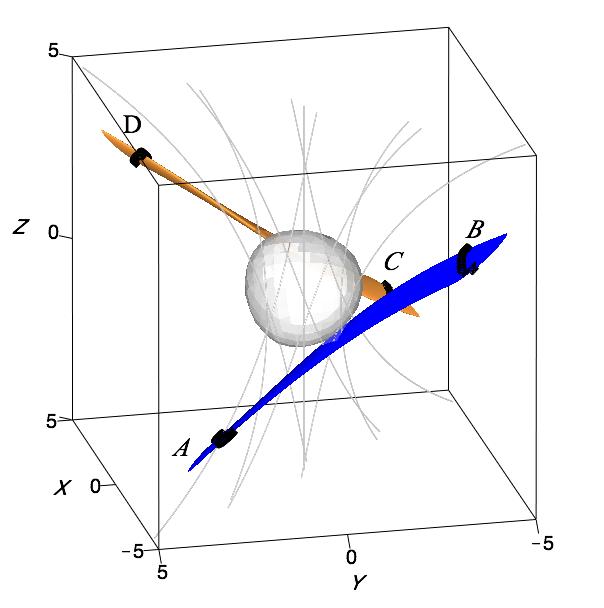}
\includegraphics[width=5.2cm]{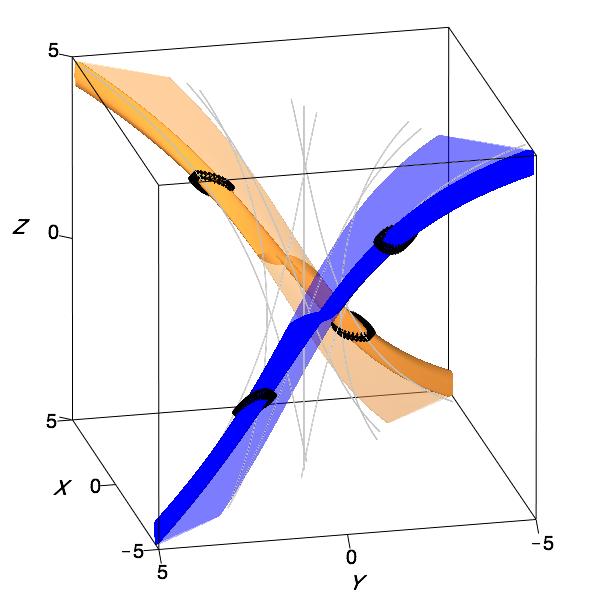}
\includegraphics[width=5.2cm]{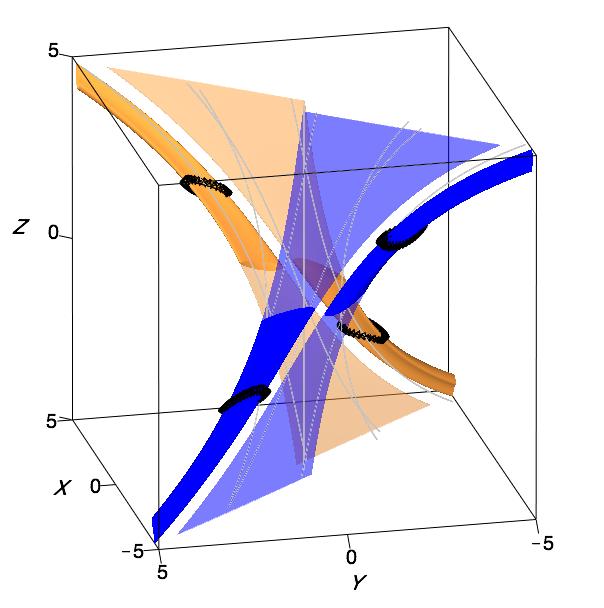}
\includegraphics[width=5.2cm]{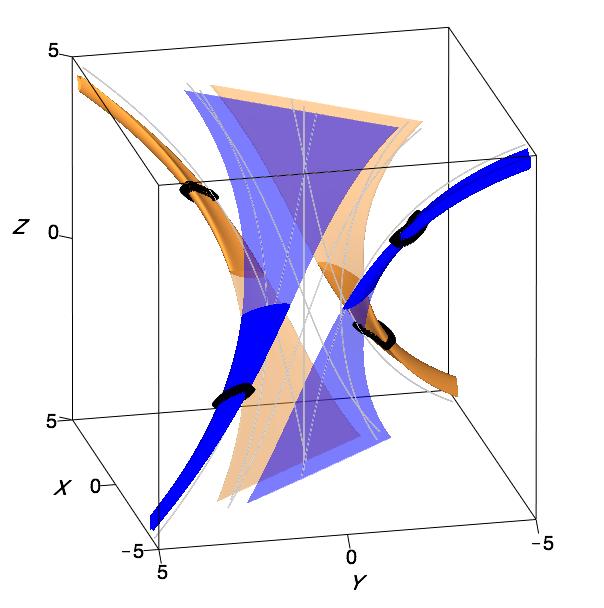}
\includegraphics[width=5.2cm]{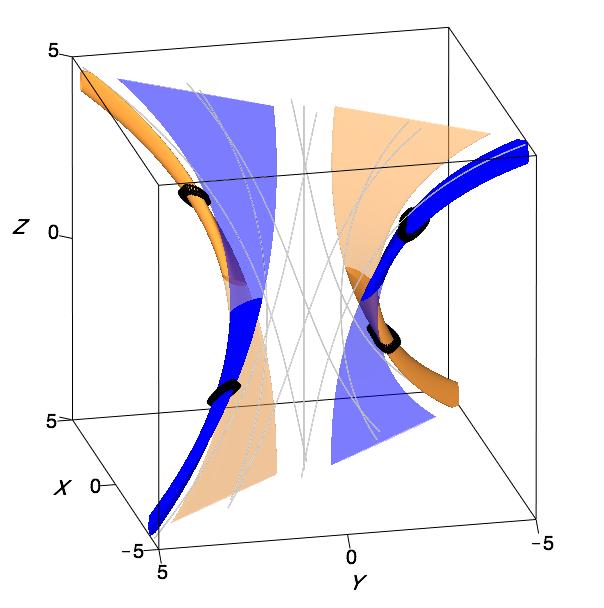}
\includegraphics[width=5.2cm]{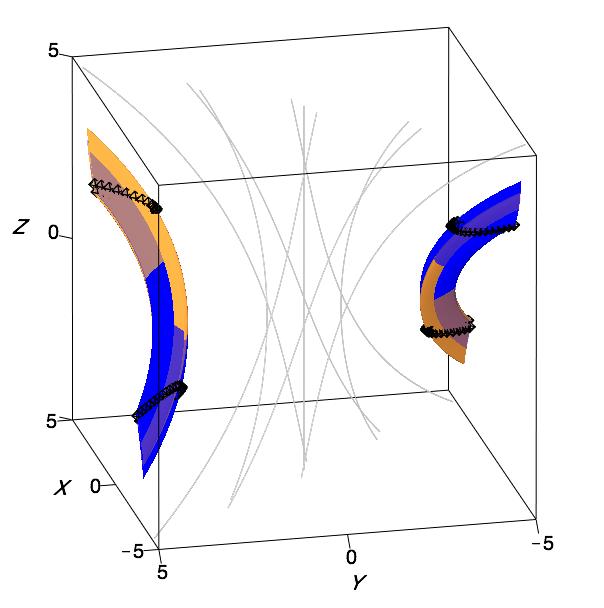}
\end{center}
\caption{Reconnection of two representative flux tubes in the magnetic field $\BB=(y, k^2 x,1)$, with $k=1.2$. The flux tubes are traced from ideal comoving footpoints (marked black), and the solid sections move at the local plasma velocity (outside $D$), while the transparent sections correspond to field lines that pass through the diffusion region.  A localised diffusion region (shaded surface in the first frame) is present around the origin. }
\label{tubeflip}
\end{figure}

Perhaps surprisingly, it turns out that none of the above properties of 2D reconnection carry over into three dimensions. In general, in the presence of a localised non-ideal region (i.e.~localised region within which $\EE\cdot\BB\neq 0$), a flux transporting velocity $\ww$ {\it does not} exist anywhere in the vicinity of the diffusion region (for a proof, see \cite{priesthornig2003}). The result is that, if one follows magnetic field lines from footpoints comoving in the ideal flow, they appear to split {\it as soon as they enter the non-ideal region}, and their connectivity changes {\it continually and continuously} as they pass through the non-ideal region (see Figure \ref{tubeflip}). In other words, between any two neighbouring times $t$ and $t+\delta t$, {\it every} field line threading the non-ideal region experiences a change in connectivity. Consequently, magnetic field lines are not reconnected in a one-to-one fashion as in 2D. To illustrate, let us  consider two field lines which are about to enter the diffusion region, one of which connects plasma elements labelled A and B, the other of which connects plasma elements labelled C and D (as in Figure \ref{tubeflip}). Then if the field lines are chosen such that after reconnection A connects to C, then B will {\it not} be connected to D. This property has profound implications for the way in which the magnetic flux is restructured by the three-dimensional reconnection process -- we can no longer think of a simple cut and paste of field line pairs. 

The above properties are demonstrated in Figure \ref{tubeflip}. Representative flux tubes are traced from four cross sections, chosen such that at the initial time they form a pair a flux tubes. The plots are based on the steady-state kinematic solution of \cite{hornig2003}, with $\BB=(y, k^2 x,1)$ and a diffusion region, $D$, localised around the origin. Note that in this solution the resistivity has been localised in order to obtain an analytical solution -- however, the above-described topological properties of the flux evolution are not dependent on this localisation, and are still present when the non-ideal region is self-consistently localised through the formation of a localised current layer (as discussed later). As the flux tubes enter the diffusion region, they immediately begin to split, with field lines from cross-sections A and B (say) no longer being coincident. In frames 2-5 of the figure, the apparent `flipping' (or `slip-running') of field lines is demonstrated. The solid sections of the flux tubes are traced from ideal comoving footpoints (marked black) and move at the local plasma velocity (outside $D$), while the transparent sections correspond to field lines traced into and beyond the diffusion region, and appear to flip past one another at a velocity that is different from the local plasma velocity (until they exit the non-ideal region). Note that while in the initial state we began with two flux tubes, in contrast to the 2D case (shown in Figure \ref{tube2d}), after reconnection (final frame, Figure \ref{tubeflip}) the four cross-sections do not match up to form two unique flux tubes. 

The above 3D reconnection properties hold regardless of the topology of the magnetic field in the vicinity of the reconnection region: for further details, the reader is referred to the papers by \cite{schindler1988,hornig2001,hornig2007a,hornig2007b,priesthornig2003}. However, there are some profound differences that do occur when different magnetic field topologies are considered, so that as a result we may consider that there are a number of different magnetic reconnection regimes in 3D.  
These regimes can be split into non-null reconnection (including reconnection in QSLs), null point reconnection (which has recently been categorised in a new way \citep{priest2009} in response to new numerical experiments), and separator reconnection.
In the following sections, we go on to consider each of these regimes in turn.

\section{3D magnetic reconnection regimes: non-null reconnection}\label{nonnullsec}

As discussed above, in three dimensions, magnetic reconnection may occur in current layers which are not associated with magnetic nulls. The continuous change of connectivity of field lines traced from comoving footpoints has led to such reconnection being termed variously {\it magnetic flipping} \citep{priest1992} or {\it slip-running reconnection} (if the virtual flipping velocity exceeds some threshold, \cite{aulanier2006}). As discussed above, there are many different mechanisms by which the current layers may form. However, some basic properties of the resulting reconnection process will be universal.

A major step in understanding the properties of 3D non-null reconnection has been made by \cite{hornig2003}. They considered the kinematic problem in which Ohm's law and Maxwell's equations are solved but the equation of motion is neglected (although it turns out that the solution solves the equation of motion in the limit of slow flows). The solution is obtained by imposing a steady-state magnetic field and plasma resistivity, and solving Ohm's law for the electric field and plasma velocity via
\begin{equation}\label{kinsol}
\Phi=\int \eta \JJ\cdot\BB \, ds, \qquad \EE=-\nabla\Phi, \qquad \vv_{\perp} = \frac{(\EE-\eta\JJ)\times\BB}{B^2},
\end{equation}
where $\vv_\perp$ is the plasma velocity perpendicular to the magnetic field -- the parallel component being arbitrary in this approximation.
In order to obtain an analytical solution, the magnetic field and resistivity profiles must then be chosen in combination such that the first equation may be integrated. 

\cite{hornig2003} chose a magnetic field consisting of a hyperbolic X-point plus a uniform field, specifically $\BB={B_0} (y,k^2 x,1)/L$. This linear magnetic field has the advantage that the field line mapping and its inverse can be expressed in closed form --  a property that makes it possible to find closed-form solutions. Since the magnetic field is linear, the current within the volume is uniform: $\JJ=(0,0,B_0(k^2-1)/L\mu_0)$. Since the authors' aim was to study an isolated 3D reconnection process -- the generic case in astrophysical plasmas -- the resistivity was chosen to be localised around the origin. The resulting electric and velocity fields are therefore fully 3D, and magnetic field lines can be traced from ideal footpoints on either side of the diffusion region, since the product $\eta\JJ$ is localised.  

\begin{figure}
\label{potdropfig}
\begin{center}
\includegraphics[width=6cm]{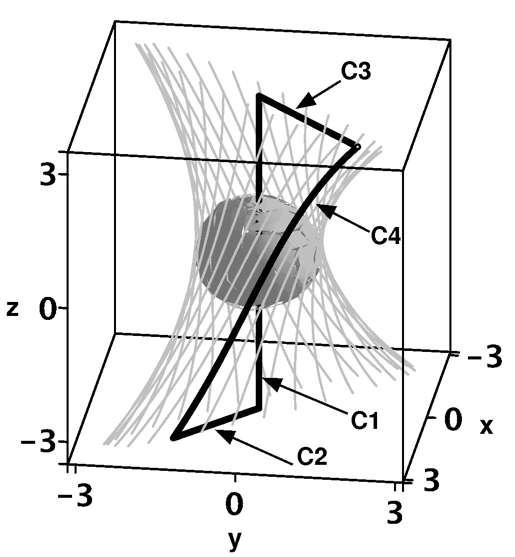}
\end{center}
\caption{Potential drop around a closed circuit for the non-null reconnection solution of \cite{hornig2003}.}
\end{figure}

Solving Equations (\ref{kinsol}), one can show that the plasma flow required to maintain this steady state configuration is a counter-rotational flow. More precisely, the flow is confined to field lines which thread the non-ideal region, with field lines above and below the non-ideal region (with respect to the direction of $\BB$) rotating in opposite senses. One can demonstrate that this is a necessary property of the solution that follows directly from the presence of a 3D-localised parallel electric field within a region of non-vanishing magnetic field. That is, this property is independent of the specific choice of spatial profiles of $\BB$ and $\eta$, or indeed the fact that $\eta$ rather than $\JJ$ is localised (a localisation of the current would be a more physically plausible way to localise $E_\|$, but is not compatible with the method of solution). This is clear if one considers the potential drop around the closed loop displayed in Figure \ref{potdropfig}, which must be zero since $\nabla\times\EE={\bf 0}$. Now, $\EE\cdot\BB\neq 0$ along the central field line and since $\EE$ does not change sign along that line  then there must be a potential drop along $C1$. Since $\EE\cdot\BB=0$ for any field line lying wholly outside the non-ideal region, there is no potential drop along $C4$. It is therefore clear from the figure that it is necessary to have non-zero potential drops along the radial lines marked $C2$ and $C3$ (of opposite signs than on $C1$). This implies a radial electric field component of opposite
sign on $C2$ and $C3$. Since this particular choice of loop is not unique, we conclude that there must be an electric field component along any such radial line, and the combination of a radial electric field with a vertical component of the magnetic field implies that there must be an azimuthal flow within this envelope of flux that threads the non-ideal region.

As a consequence of the counter-rotational flows, field lines followed from the ideal region above and below the non-ideal region appear to undergo a `rotational slippage' with respect to one another. This rotational slippage is quantified by the reconnection rate calculated in Equation (\ref{recratedef}). It is worth emphasising that this characteristic flow structure for 3D non-null reconnection is very different to the classical 2D reconnection picture in which the characteristic flow structure is of stagnation type. The counter-rotational flows are a signature of the helicity production (decay) in 3D reconnection: if one writes down an evolution equation for the magnetic helicity then $\EE\cdot\BB$  appears as a source term. The solution of \cite{hornig2003} allows for the addition of an ideal flow via the constant function of integration in the integral in Equation (\ref{kinsol}). This function must be independent of $s$, i.e.~constant along field lines, but may vary from one field line to another. The authors considered the effect of adding a flow with a hyperbolic structure in the $xy$-plane to transport magnetic flux into and out of the diffusion region (by adding a term proportional to $x_0y_0$ to $\Phi$, where $(x_0,y_0)$ is the point of intersection of a field line with the $z=0$ plane). The result is that field lines are brought into the non-ideal region, are split apart by the counter-rotational flows, and exit differently connected in opposite quadrants of the flow. The evolution of a particular pair of flux tubes for one of these solutions is shown in Figure \ref{tubeflip}.

One should note that the idea of a localised region of non-zero $E_\|$ being associated with a rotation in the plane perpendicular to the field has been described before by e.g.~\cite{hesse1991}. The above solution has been refined and put on a firm footing by solving the full system of MHD equations using an expansion scheme \citep{wilmotsmith2006,wilmotsmith2009c,alsalti2009}, and the properties of the solution were also verified in a resistive MHD simulation \citep{pontingalsgaard2005}. The solution describes the properties of a generic 3D reconnection process in a steady-state magnetic field in the absence of null points. In recent years various numerical simulations have explored the effects of 3D reconnection in more or less complicated magnetic field configurations. Notably, \cite{linton2001} and \cite{linton2003} have investigated the interaction of magnetic flux tubes in an otherwise field-free environment, and discovered different possible interactions (`merge', `bounce', `tunnel' and `slingshot') depending on the relative orientations of the tubes. In addition, recent simulations have shown that the local structure of the reconnection site during a non-null reconnection process need not necessarily be hyperbolic, but may also be elliptic \citep[e.g.][]{wilmotsmith2010}.

\section{3D magnetic reconnection regimes: null point reconnection}\label{nullsec}
\subsection{Kinematic models}
As discussed above, 3D magnetic null points have been proposed as possible sites of magnetic reconnection, due to the singularities that appear there in ideal kinematic models.
The tendency of 3D null point structures to collapse (in the same way as 2D X-points) to generate currents locally has been studied by a number of authors \citep[e.g.][]{klapper1996,bulanov1997,mellor2003}. Furthermore, investigations by \cite{pontincraig2005} suggest that current singularities are a natural consequence of an ideal MHD evolution in the vicinity of a line-tied 3D null. 

Early models for 3D null point reconnection were proposed by \cite{priest1996} who considered  the ideal kinematic limit and a current-free magnetic null. However, \cite{pontin2004,pontinhornig2005} showed that the possible magnetic flux evolutions are very different when a localised diffusion region is included around the null point. They performed a similar kinematic analysis to that of \cite{hornig2003}, again imposing a steady-state magnetic field and resistivity profile, and solving for the corresponding electric field and plasma flow. It was found that the nature of the magnetic reconnection is crucially dependent on the orientation of the electric current at the null point. 

\begin{figure}
\begin{center}
\includegraphics[width=5.2cm]{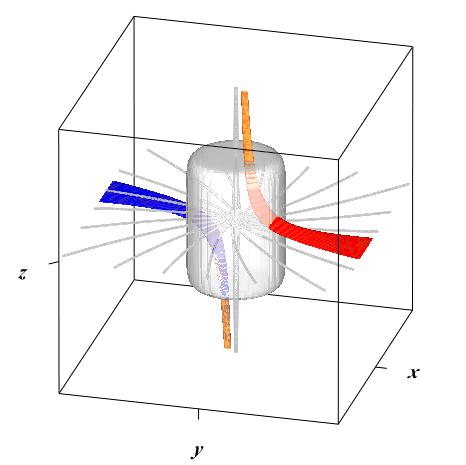}
\includegraphics[width=5.2cm]{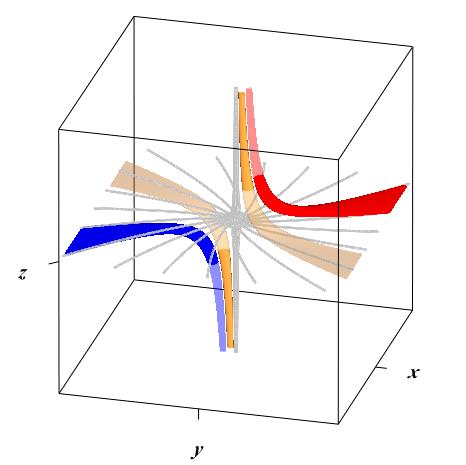}\\
\includegraphics[width=5.2cm]{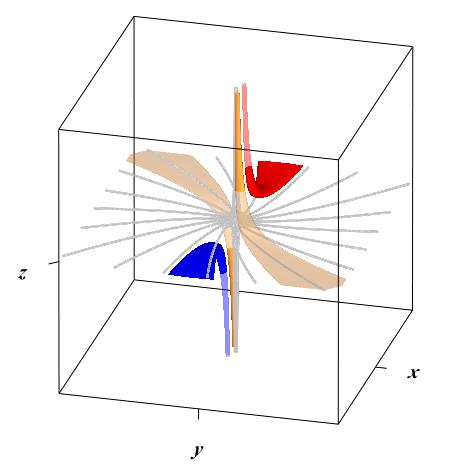}
\includegraphics[width=5.2cm]{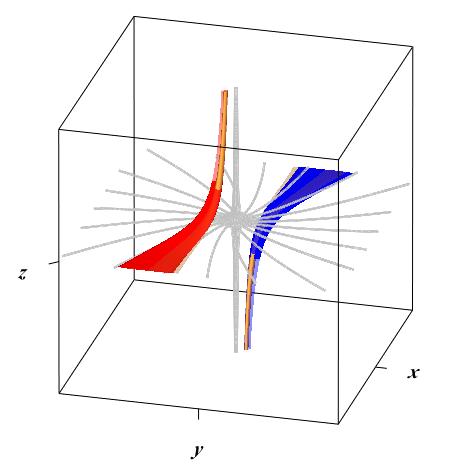}
\end{center}
\caption{Reconnection of two representative flux tubes in the magnetic field $\BB=(r, jr/2,-2z)$ in cylindrical polar coordinates, corresponding to current directed parallel to the spine (with $j=1$). A localised diffusion region is present around the null point, shown by the shaded surface in the first frame. Flux tubes are traced from four ideal comoving footpoints, with their extensions that pass through the diffusion region rendered as transparent, as in Figure \ref{tubeflip}.}
\label{tube_spinealigned}
\end{figure}

\begin{figure}
\begin{center}
\includegraphics[width=8cm]{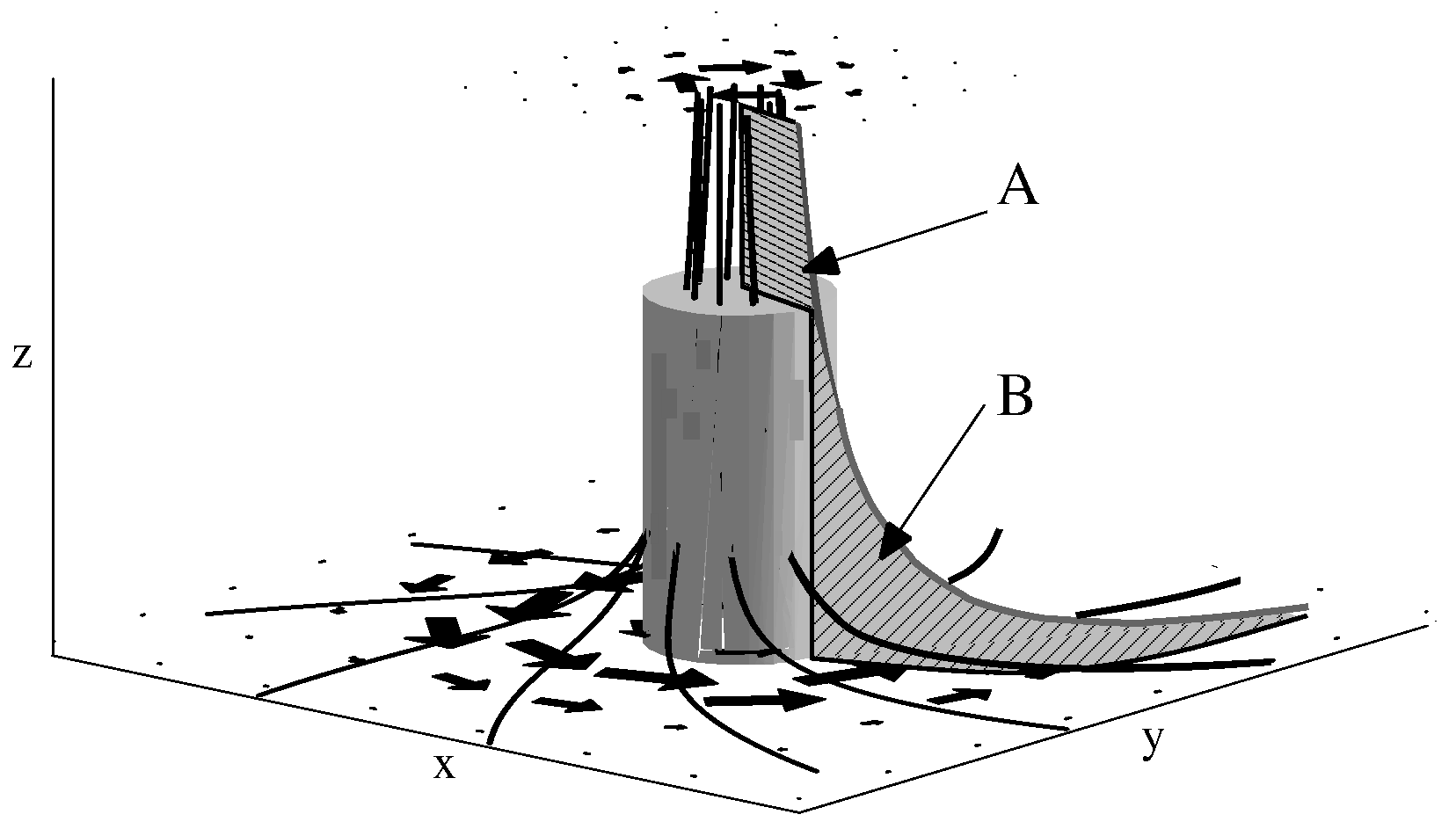}
\end{center}
\caption{Interpretation of the reconnection rate for null point reconnection with a spatially-uniform spine-aligned current (kinematic solution with $\BB=(r,jr/2,-2z)$ in polar coordinates, $j=1$). Heavy black lines are magnetic field lines, the grey cylinder represents the diffusion region, and the arrows show the plasma flow in two representative planes of $z=const$. Magnetic flux is transported through surfaces `A' and `B' at different rates.}
\label{spinerecrate}
\end{figure}

If the current is directed parallel to the spine of the null, then there are counter-rotational flows, centred on the spine \citep{pontin2004,wyper2010}. The change of connectivity that results from the reconnection process therefore takes the form of a rotational slippage similar to that discovered in the non-null case described above (see Figure \ref{tube_spinealigned}). Importantly, there is no flux transport across either the spine or fan. In the simple model of \cite{pontin2004} the reconnection rate is given by the integrated parallel electric field along the spine (by symmetry). The reconnection rate so determined quantifies the rotational slippage. Specifically, it measures the difference between the rate of (rotational) flux transport in the ideal region on either side of the diffusion region - i.e.~the difference in the rate of flux transport through the two surfaces (`A' and `B') shown in Figure \ref{spinerecrate}.

\begin{figure}
\begin{center}
\includegraphics[width=5.2cm]{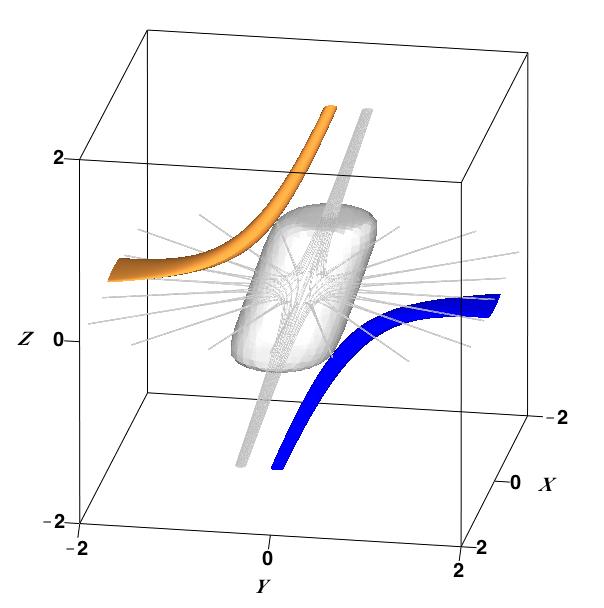}
\includegraphics[width=5.2cm]{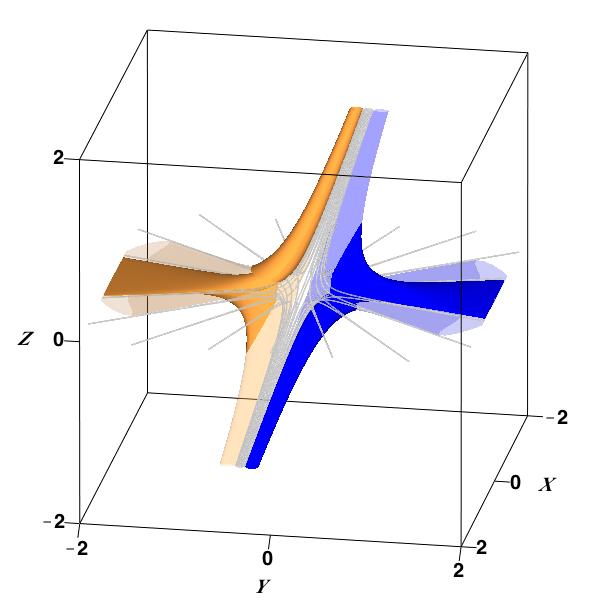}\\
\includegraphics[width=5.2cm]{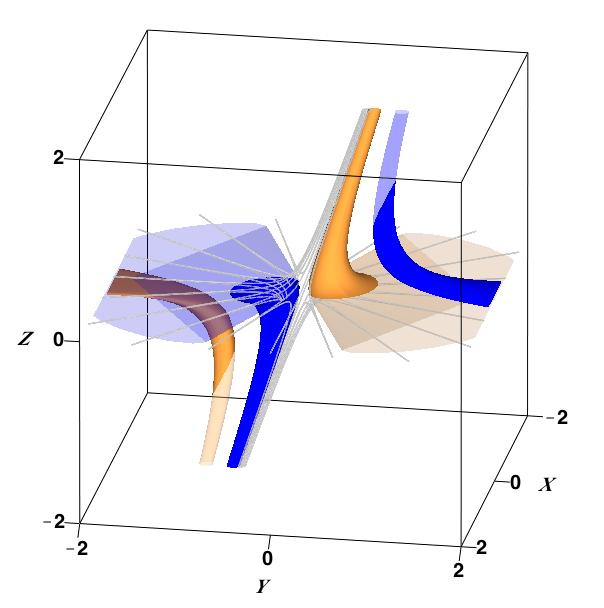}
\includegraphics[width=5.2cm]{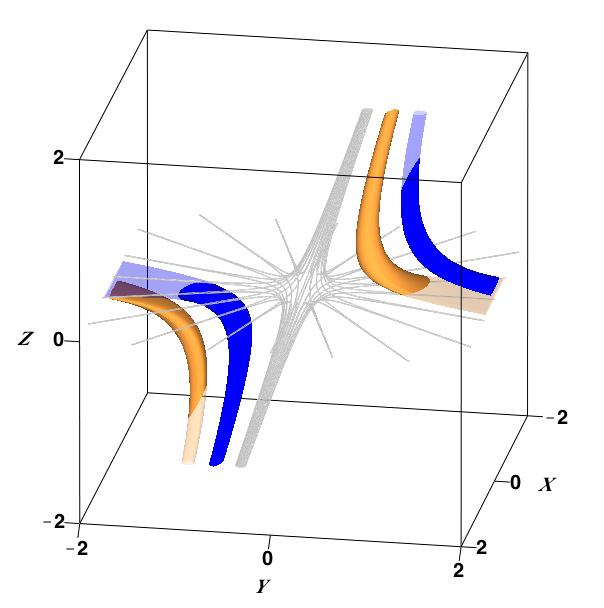}
\end{center}
\caption{Reconnection of two representative flux tubes in the magnetic field $\BB=(x,y-jz,-2z)$, corresponding to current directed parallel to the fan plane (with $j=1$). A localised diffusion region is present around the null point, shown by the shaded surface in the first frame. Flux tubes are traced from four ideal comoving footpoints, with their extensions that pass through the diffusion region rendered as transparent, as in Figure \ref{tubeflip}.}
\label{tube_fanaligned}
\end{figure}

By contrast, when the current is directed parallel to the fan surface (and is non-zero at the null itself), plasma flows cross both the spine and fan of the null, transporting flux both through/around the spine line, and across the fan separatrix surface \citep{pontinhornig2005}, as shown in Figure \ref{tube_fanaligned}. In this case, the reconnection rate can be shown to quantify the rate at which magnetic flux is transported across the separatrix surface in the ideal region---an interpretation that more closely resembles the two-dimensional picture.

The above described solutions suggest two main modes of magnetic reconnection at 3D nulls. However, in these kinematic models the diffusion region was artificially localised. The question still remains as to what types of current concentrations form self-consistently at 3D nulls in the dynamic regime. This has been investigated in a series of numerical simulations \citep{rickard1996,galsgaard2003,pontingalsgaard2007,pontinbhat2007a}. The results have led \cite{priest2009} to propose a new categorisation of 3D null point reconnection regimes, as follows.

\subsection{Torsional spine and fan reconnection}
\cite{rickard1996} and \cite{pontingalsgaard2007} investigated the propagation of disturbances towards symmetric 3D null points (where the fan eigenvalues are equal). In both studies, a general disturbance was decomposed into rotations (in planes perpendicular to the spine) and shearing motions. Rotational motions were found to behave in an essentially Alfv{\' e}nic manner: they propagate along field lines, and accumulate around the spine line or fan plane. The locations of the spine and fan themselves remain undisturbed from their orthogonal potential configuration. These results are analogous to the properties of Alfv{\' e}n wave propagation towards 2D X-points, summarised by \cite{mclaughlin2010}. In each case, due to the hyperbolic geometry of the magnetic field, the current intensifies as the length scales perpendicular to the spine or fan become shorter. This intensification ceases once these length scales become sufficiently short that diffusion becomes important.

\begin{figure}
\begin{center}
\includegraphics[width=6cm]{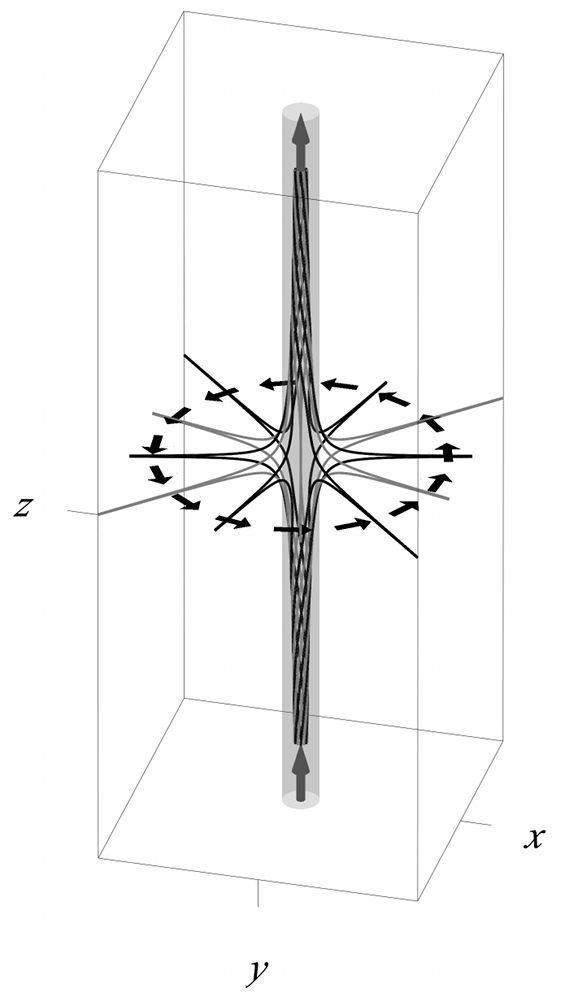}
\end{center}
\caption{Schematic diagram of torsional spine reconnection at an isolated null. Black and grey lines are magnetic field lines, the shaded surface is a current density isosurface, the grey arrows indicate the direction of the current flow, while the black arrows indicate the driving plasma  velocity.}
\label{torspine}
\end{figure}

{\it Torsional spine reconnection} occurs in response to a rotational disturbance of the fan plane. The disturbance propagates to the spine, around which an extended tube of current forms. This current tube is generated by a twisting of the magnetic field lines locally around the spine line, and as such the current vector is directed parallel to the axis of the tube, i.e.~parallel to the spine, see Figure \ref{torspine}. Due to the orientation of the current, the magnetic reconnection that occurs within the current layer takes the form of a rotational slippage, as discussed above. In a continuously driven system, a quasi-steady state will be reached when the rotational advection that increases the twist of the magnetic field around the spine -- and thus the current -- is balanced by this rotational slippage.

\begin{figure}
\begin{center}
\includegraphics[width=6cm]{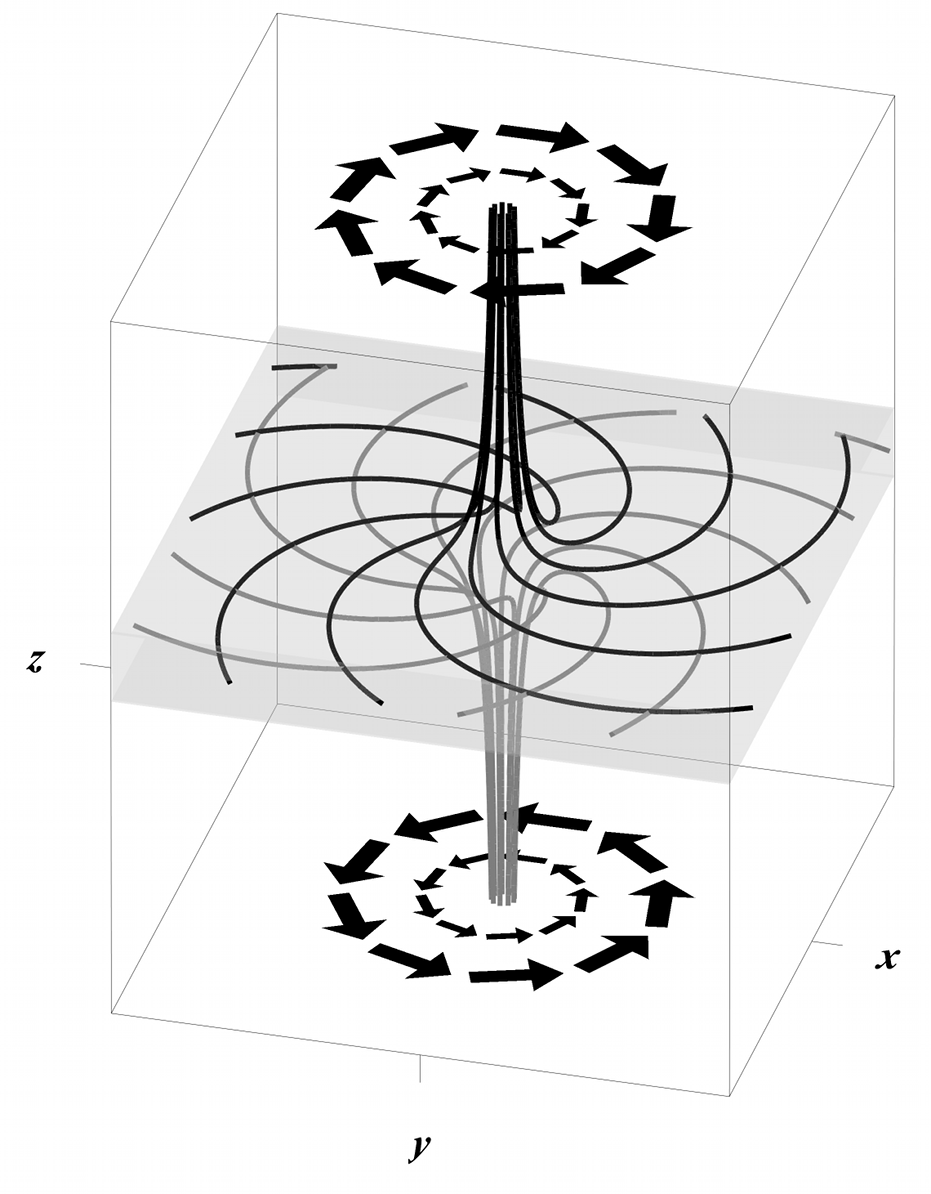}
\end{center}
\caption{Schematic diagram of torsional fan reconnection at an isolated null. Black and grey lines are magnetic field lines, the shaded surface is a current density isosurface, and the black arrows indicate the driving plasma  velocity.}
\label{torfan}
\end{figure}

{\it Torsional fan reconnection} occurs in response to a rotational disturbance around the spine. The perturbation propagates as a helical Alfv{\' e}n wave towards the fan, where a planar current layer develops \citep{galsgaard2003}, as shown in Figure \ref{torfan}. While away from the null the current is dominated by its components parallel to this plane, it flows through the null parallel to the spine. Therefore a rotational slippage of magnetic flux is expected from the kinematic models, and indeed this is the form that the reconnection takes. Again, a quasi-steady current layer would be expected to form once the increased twisting of field lines is balanced by this rotational slippage.

\subsection{Spine-fan reconnection}
\begin{figure}
\begin{center}
\includegraphics[width=8cm]{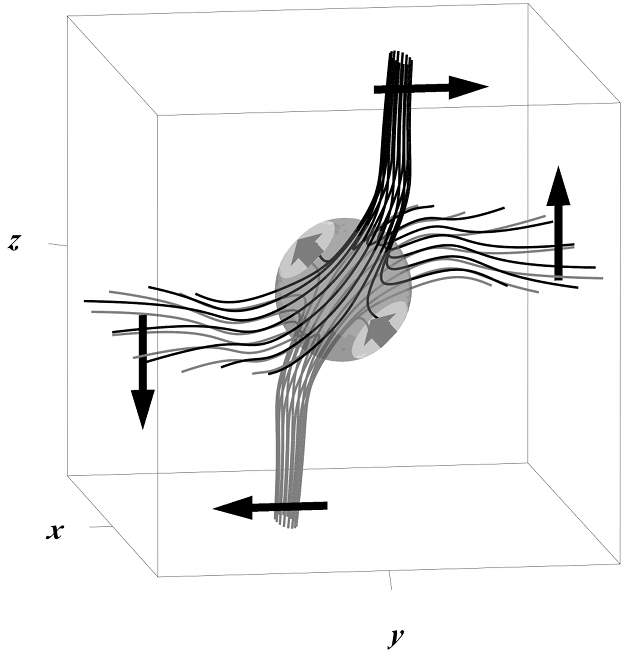}
\end{center}
\caption{Schematic diagram of spine-fan reconnection at an isolated null. Black and grey lines are magnetic field lines, the shaded surface is a current density isosurface, the grey arrows indicate the direction of the current flow, while the black arrows indicate the driving plasma  velocity.}
\label{spinefan}
\end{figure}

The torsional spine and torsional fan reconnection modes discussed above require a rather organised rotational driving motion -- and it is thus anticipated that the most common regime of reconnection to occur at 3D nulls is the {\it spine-fan reconnection} mode. This mode of reconnection occurs within a current sheet that is localised in all three dimensions around the null. Such a current concentration is found to form when a shear disturbance of {\it either} the spine {\it or} the fan occurs \citep{pontinbhat2007a}. Since in this case the disturbance propagates {\it across} magnetic field lines as it localises at the null, its behaviour has the properties of a magnetoacoustic wave. The current layer at the null is formed by a local collapse of the magnetic field -- the spine and fan collapse towards one another, with the current sheet locally spanning them both, as depicted in Figure \ref{spinefan}. The plane in which the spine and fan collapse is selected by the plane of the shear disturbance, and in this plane the spine, fan, and current layer together form a Y-type structure. The current flows through the null perpendicular to this shear plane, and thus parallel to the fan surface. As the null point collapses, magnetic flux is transported through {\it both} the spine {\it and} the fan, as predicted by the kinematic model due to the current orientation. Although the majority of previous studies considered the perturbation of a symmetric null point, \cite{alhachami2010} have demonstrated that when a generic non-symmetric null is considered, while the qualitative properties of the reconnection process are preserved, the dimensions of the diffusion region and reconnection rate can vary strongly.

It is worth noting the relation of the spine-fan reconnection regime to the steady-state mathematical models proposed by \cite{craigetal1995,craig1996}. These are exact solutions of the incompressible MHD equations, and are often termed `reconnective annihilation' models, since they involve current layers that extend to infinity along either the spine or the fan. The solutions are constructed by super-imposing 1D or 2D disturbances consisting of infinite, straight field onto a background potential null. The solutions involving a planar current layer in the fan have been demonstrated to be dynamically accessible \citep{craig1998}. However, in order to maintain the planar nature of the current layer (which is imposed by the restrictive but necessary choice of low-dimensionality disturbance fields), a large pressure gradient is required within the current sheet. It has been shown in MHD simulations that when the incompressibility condition is relaxed, the pressure gradient is not able to balance the Lorentz force acting within the current layer, whereupon the magnetic field collapses to form a localised current layer at the null as described above \citep{pontinbhat2007b}. On the other hand, indications are that the spine reconnective annihilation models are not dynamically accessible \citep{titov2004,pontinbhat2007b}.

\section{3D magnetic reconnection regimes: separator reconnection}\label{sepsec}
\begin{figure}
\begin{center}
\includegraphics[width=7cm]{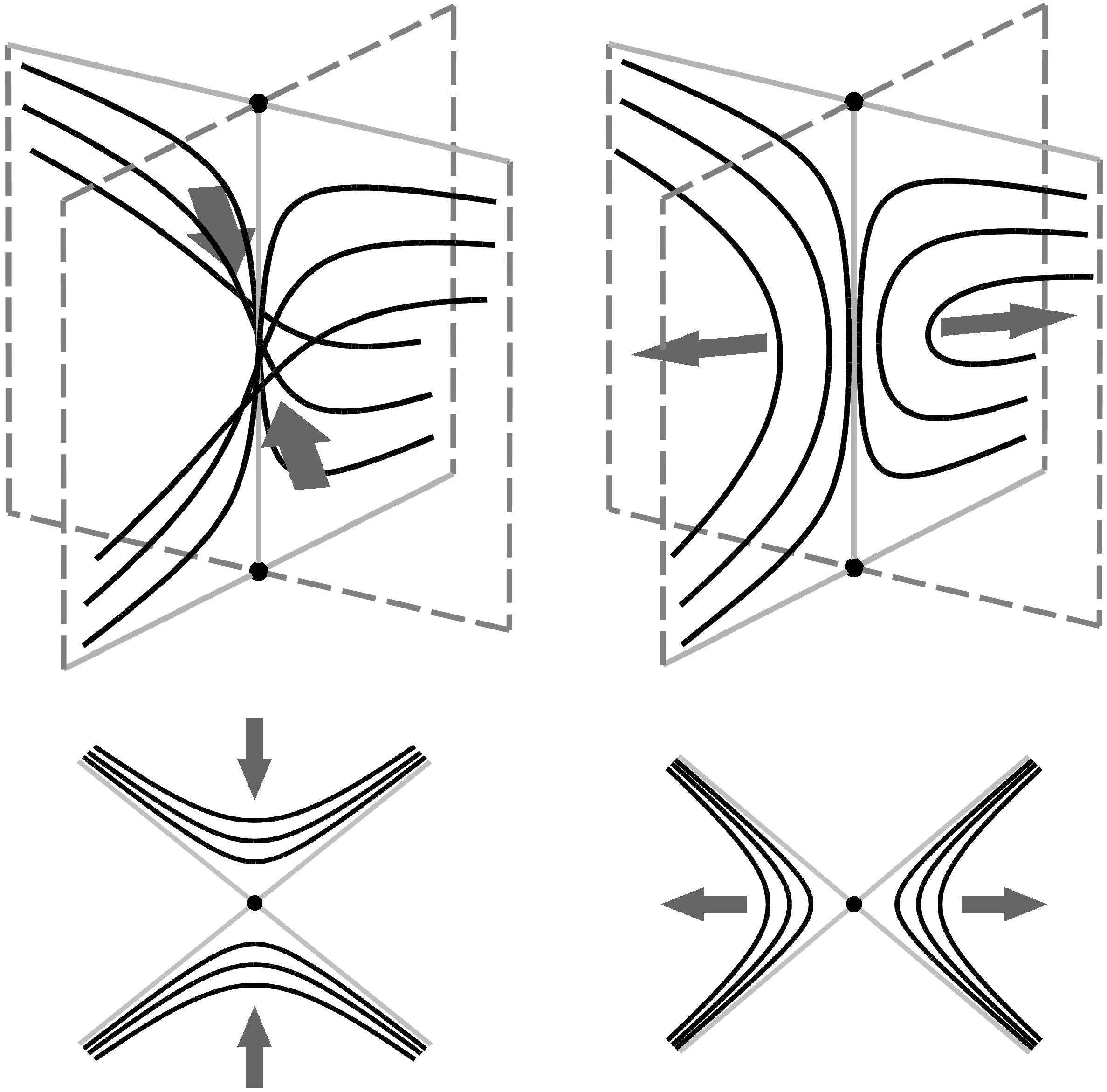}
\end{center}
\caption{Cartoon showing field lines before (left) and after (right) reconnection at a separator in the simplified kinematic picture. Black circles mark the nulls and arrows indicate the direction of plasma / field line motion. Lower images show the view looking along the direction of the separator from above.}
\label{sepcart}
\end{figure}

In addition to reconnection at isolated 3D null points and in their absence, reconnection may also occur in a configuration containing multiple magnetic nulls connected by one or more separator field lines. The form of current layers at such separator field lines has been investigated by \cite{longcopecowley1996}, while \cite{longcope1996} has proposed that currents will naturally focus at separator lines during relaxation processes in the solar corona. Early kinematic models in current-free magnetic fields predicted that separator reconnection would involve a simple cut-and-paste of field line pairs at the separator line \citep{lau1990,priest1996}. In the absence of any current, the magnetic field in a plane orthogonal to the separator has a perpendicular X-type structure, and the reconnection was envisaged -- projected on such a 2D plane -- as being much like 2D X-point reconnection, as depicted in Figure \ref{sepcart}. Such a picture was qualitatively backed up in numerical simulations by \cite{galsgaardpriest2000}. However, new results throw serious doubts on these simplified pictures.

\begin{figure}
\begin{center}
\includegraphics[width=5cm]{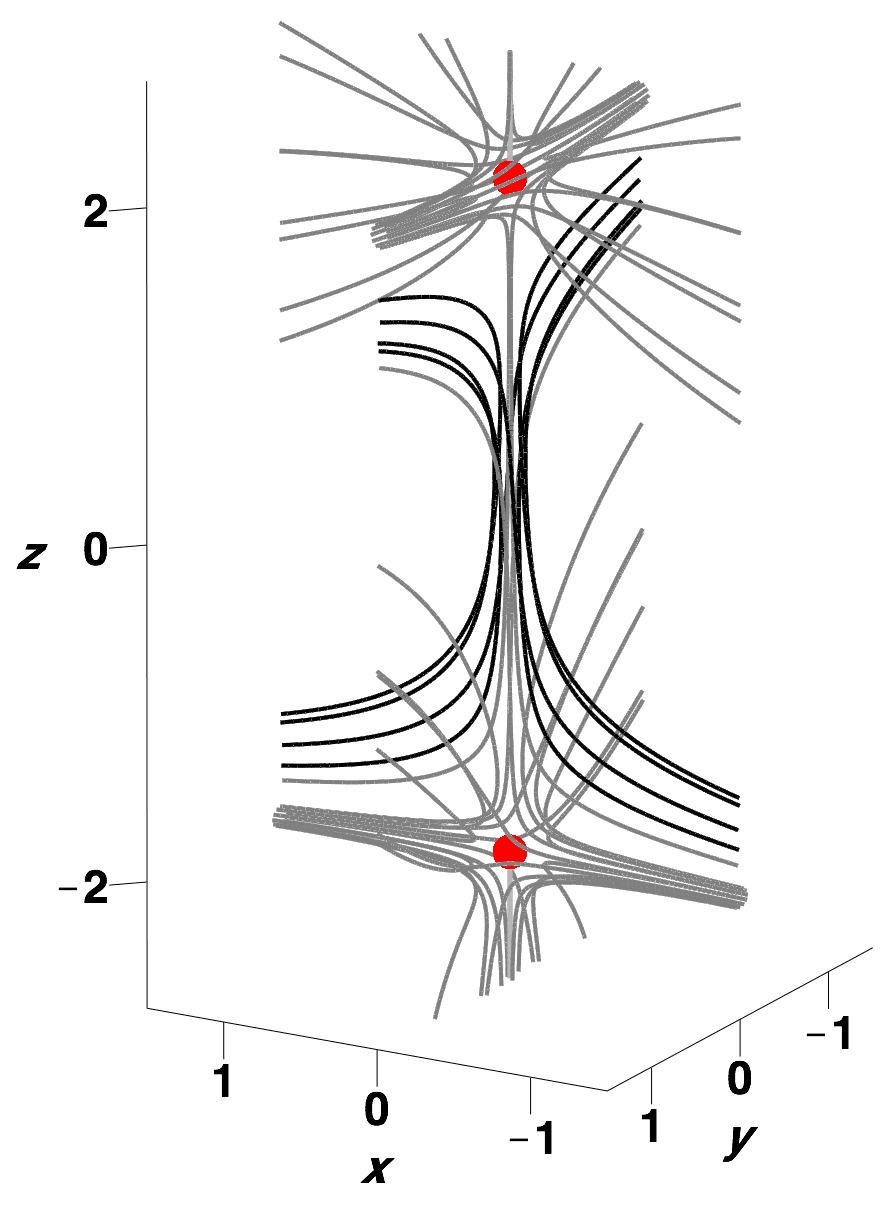}
\includegraphics[width=5cm]{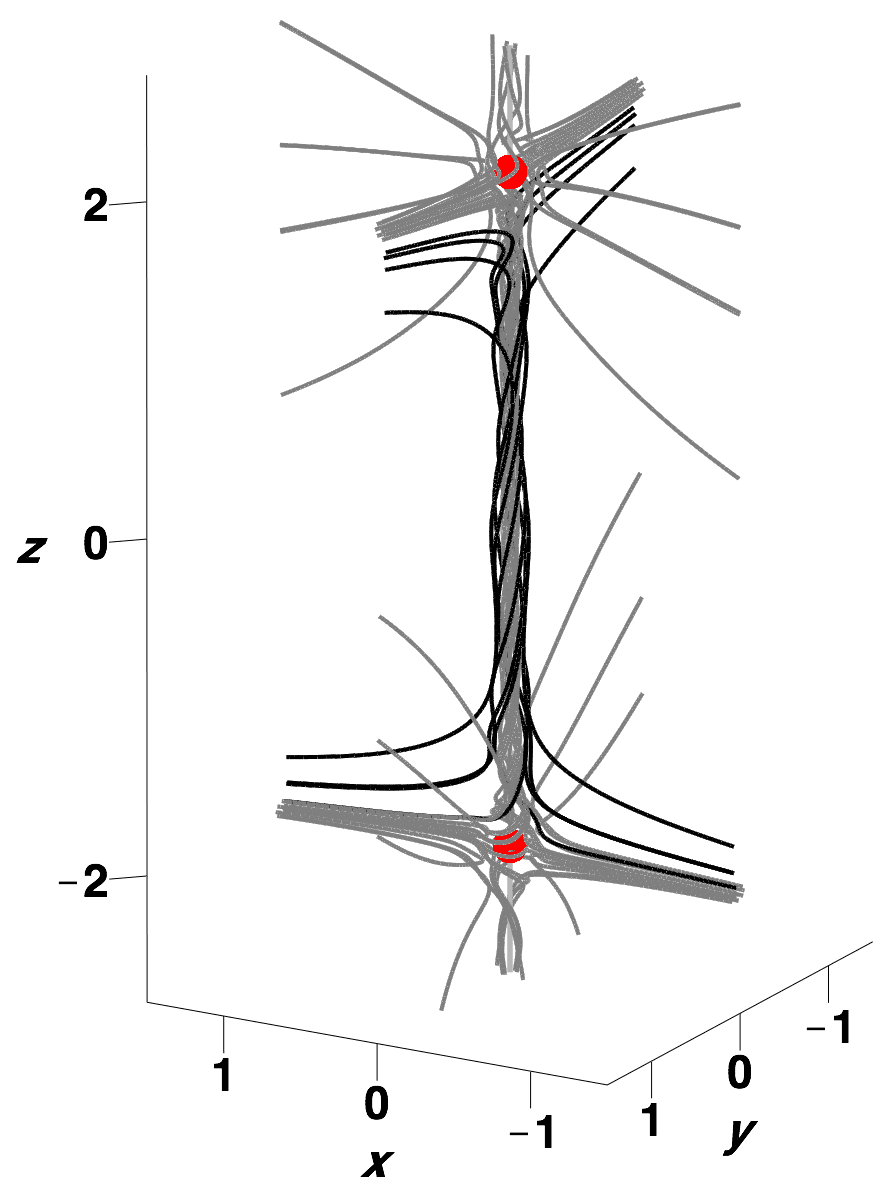}
\end{center}
\caption{Field lines for the magnetic field defined by Equation (\ref{sepeq}), with $z_0=2$ and (a) $j=0.1$ and (b) $j=0.4$. The separator runs along the $z$-axis between the two disks which represent the nulls (red online). Grey field lines are traced from footpoints near the spines of the nulls, while black field lines are traced from a circle around the separator, centred at the origin in the $z=0$ plane.}
\label{sepstruc}
\end{figure}

Separator reconnection is perhaps the least well-understood 3D reconnection regime. One major unknown property is the typical distribution of the current along the separator, and therefore whether separator reconnection shares more in common with the null or non-null reconnection regimes. Reconnective annihilation models for incompressible plasmas suggest that reconnection may be focussed either at the nulls or between them along the separator \citep{craig1999,pontin2006}. One thing that is clear is that the picture of cut-and-paste one-to-one rejoining of field lines at the separator line is over-simplified. When a localised current layer forms around a separator, the reconnection within the associated diffusion region must conform to the properties described in Section \ref{recpropsec}. Therefore there will be a continuous reconnection of field lines within the volume surrounding the separator. Indeed, as demonstrated by \cite{parnell2010}, the structure of the magnetic field in the vicinity of the diffusion region may be significantly more complex than previously expected. The authors investigated in detail the evolution of the magnetic field in a simulation in which two patches of opposite polarity magnetic flux on the boundary were driven past one another \citep[see also][]{haynes2007}. While there were only two isolated, unconnected nulls in the initial configuration,  during the evolution a number of separators were formed, around which the current was found to be focussed. 
Although a separator must have a hyperbolic field (in the plane locally perpendicular to the field) near the nulls, it can have hyperbolic as well as elliptic structure away from the nulls.  In the simulation,  changes of the structure both along the separator as well as in time were found.
It is natural that an elliptic field such as this be present in the vicinity of a strong current along the separator. This is demonstrated in Figure \ref{sepstruc}. Field lines are plotted for a magnetic field consisting of a potential component defining a separator structure plus a component that defines a line current. Specifically, the magnetic field is
\begin{eqnarray}
\BB&=&\left(x\left(z-3z_0\right)\ , \ y\left( z+3z_0\right) \ , \ z_0^2-z^2 +\frac{1}{2}(x^2+y^2) \right) \nonumber\\
& &+ j\left( -40y e^{-20x^2-20y^2} \ , \ 40x e^{-20x^2-20y^2} \ , \ 0  \right) \label{sepeq}
\end{eqnarray}
which contains nulls at $x=y=0$, $z=\pm z_0$, with a separator located at $-z_0<z<z_0$. Part (a) shows the case of a relatively weak current, $j=0.1$, for which the magnetic field around the separator is hyperbolic. However, when the current is increased ($j=1$, shown in part (b)), the field lines spiral around the separator.
Clearly, reconnection at a separator with elliptic local magnetic field must be rather different to the simple early models. Indeed, \cite{parnell2010} identified the presence of counter-rotating flows around the separator on either side of localised enhancements in the parallel electric field, a signature of the non-null reconnection described in Section \ref{nonnullsec}.

\section{3D magnetic reconnection in simulations and observations}\label{obssec}

Magnetic reconnection is a key ingredient for many astrophysical processes, from the collapse of accretion disks through  to stellar dynamos and flares. 
Identification of the 3D magnetic reconnection regimes described above has assisted with the diagnosis of results from -- and been motivated by --  observations and numerical simulations. Those observations and simulations that invoke specific 3D reconnection models tend to come from the study of the Solar atmosphere and Earth's magnetosphere, as comparatively well-resolved data is available for these environments. 

3D reconnection in the absence of null points is seen, for example, in 3D simulations of the Parker coronal heating scenario \citep[e.g.][]{hendrix1996,galsgaard1996,rappazzo2008}. While the topology of the magnetic field in the vicinity of the reconnection sites is not determined in these studies, the presence of a strong background `guide' field ensures the absence of topological features within the domain.  In the solar atmosphere, observed sites of energy release are often taken as signatures of a local magnetic reconnection process. While the magnetic field in the corona cannot be directly measured, many studies have  attempted to determine the local magnetic field structure around the reconnection site by extrapolation from vector magnetograms. A number of these studies find the locations of energy release to be well-correlated with the locations of quasi-separatrix layers within the coronal volume \citep[e.g.][]{demoulin1994,demoulin1997,mandrini2006,titov2008}. Furthermore, numerical modelling of the loss of stability and subsequent eruption of flux ropes has implicated reconnection in a QSL beneath the flux rope \citep{titov1999,kliem2004,titov2008}. A new technique for quantifying the reconnection rate in such simulations using so-called `slip-squashing factors' has recently been proposed \citep{titov2009}. Observations of the solar corona give strong indications that the continuous change of connectivity associated with 3D reconnection truly occurs there. \cite{aulanier2007} reported observations by the X-ray telescope onboard the {\it Hinode} satellite of  slippage of coronal loops which implicated non-null (slip-running) reconnection in a QSL. Further evidence was presented by \cite{masson2009} who described {\it TRACE (Transition Region And Coronal Explorer)} observations of propagating bright sources along a flare ribbon associated with the fan surface of a coronal null point, related to the flipping of field lines during spine-fan reconnection at the null.

Extrapolations of magnetic fields in solar active regions indicate that different magnetic reconnection modes are important for different flare events. In particular, while the references above found no 3D magnetic nulls in the vicinity of energy release events, there are also a number of observations which suggest that null points are a common feature in flaring or eruptive locations in active regions and the quiet sun \citep[e.g.][]{fletcher2001,mandrini2006,luoni2007,UgarteUrra2007,torok2009,masson2009}. Furthermore, while nulls are by no means found to be associated with all solar eruptions, a statistical study by \cite{barnes2007} showed that active regions containing nulls are more susceptible to yield eruptions. 3D null point reconnection, via the spine-fan reconnection mode, has been proposed as a mechanism for polar jets \citep{pariat2009} as well as for CMEs via the `magnetic breakout' model \citep{antiochos1999,Lynch2008}.

The hypothesis that current accumulates naturally at separator lines has been backed up by various numerical simulations \citep{galsgaard1997,galsgaardreddy1997,haynes2007}. \cite{longcope2005a} have inferred the presence of separator reconnection in the corona based on observations of the emergence of a new active region in the vicinity of a pre-existing active region. One of the great difficulties in diagnosing the reconnection mechanism in 3D numerical simulations has been to determine the topology of the magnetic field in the vicinity of the reconnection site. However, great progress has recently been made with the development of new algorithms to effectively determine the topological `skeleton' of complex 3D magnetic fields \citep{haynes2007b,haynes2010}.
While many observational signatures of reconnection in the Earth's magnetosphere have been treated as quasi-2D and interpreted as such, it is now clear that 3D topology is also of crucial importance there. Examining the magnetic topology of global magnetospheric simulations (with Northward IMF, clock angle $45^\circ$), \cite{dorelli2007} discovered clusters of nulls in the cusp regions connected on the dayside by a separator line. Furthermore, the presence of multiple 3D magnetic nulls in the tail current sheet connected by webs of separators has been inferred from {\it in-situ} observations made by the Cluster spacecraft \citep{xiao2007,deng2009}, suggesting that some kind of complex 3D tearing-like process may be occurring there.

\section{Summary and outlook}\label{sumsec}
Magnetic reconnection is a universal process in astrophysical plasmas. It facilitates the release of stored magnetic energy by permitting changes of magnetic topology (or just changes of field line connectivity for reconnection in the absence of topological structures) and as such is a key ingredient of many energetic processes in these environments. It is only in recent years that the rich geometrical and topological structure of these astrophysical plasmas has begun to be appreciated, following great advances in observations. 

In 3D, the qualitative properties of reconnection are much richer than in 2D, where reconnection occurs only at magnetic X-type null points. There are many topological and geometrical features of a magnetic field that may be favourable sites for current growth. 3D reconnection processes in different magnetic field structures have different characteristic properties, and as such can be classified into separate regimes. Reconnection may occur in the absence of any null points, with the presence of counter-rotational flows on either side of the diffusion region being a characteristic feature. Reconnection may also occur at isolated null points, with `torsional spine', `torsional fan' and `spine-fan' reconnection modes thus far identified. The most common of these -- spine-fan reconnection -- involves a local
collapse of the null to form a current sheet (focussed at the null) that locally spans both the spine and fan, with
flux transfer through spine line and fan (separatrix) surface. Furthermore, reconnection may occur at separator lines connecting pairs of nulls, in which the properties of the reconnection process may share some properties with null or non-null reconnection modes. 
While each of the reconnection modes has its own characteristics, they all share some fundamental properties that make them distinct to 2D ($\EE\cdot\BB=0$) reconnection. In particular, the non-existence of a unique flux-conserving velocity anywhere within the diffusion region (i.e.~for any field lines threading the diffusion region), which implies that reconnection occurs throughout the diffusion region, not at a single point as in 2D. As a result, there is no one-to-one cut-and-paste rejoining of field lines.

While we are now beginning to understand some of the properties of 3D reconnection, there is much left to discover. Future advances are likely to be led by  large-scale MHD simulations being developed alongside fundamental theory. 
Many important open questions remain, such as:
\begin{itemize}
\item
What are the quantitative properties of the different 3D reconnection regimes? Thus far the vast majority of our knowledge is of a qualitative nature, and quantitative studies are required to probe, for example,  the diffusion region dimensions. What is the range of possible reconnection rates in each regime, and what determines the reconnection rate? How do these numbers scale with different plasma parameters? While 3D reconnection studies have so far (by necessity) used the MHD approximation, it will be important in the future to investigate the implications of including additional physics such as Hall and electron pressure tensor electric fields in Ohm's law, to better model the physics within the diffusion region.
\item
What is the relationship between null, non-null, and separator reconnection? 
\item
Which of these modes of reconnection is most important in more realistic (complex) magnetic fields -- when the local models discussed above are embedded within global magnetic field configurations? One step towards answering this question would be to determine the observational signatures of the different reconnection regimes.
\item
What are the most common mechanisms of current sheet formation in complex 3D magnetic fields?
How does the reconnection influence the global evolution of the magnetic field? (This may be more complex than previously appreciated -- recent MHD simulations have shown that the distribution of reconnection sites may be highly complex and flux may be reconnected multiple times in generic 3D MHD evolutions \cite{parnell2008,pontin2010})
\item
How is the magnetic energy transferred to other forms, and ultimately dissipated, in 3D reconnection?
This includes the important question of the pattern and efficiency of particle acceleration in each 3D reconnection regime.
\end{itemize}

\section{Acknowledgements}
This review is based on a paper presented at the 38th COSPAR assembly. The author would like to thank the organisers for the invitation to present the review, and gratefully acknowledges financial support from COSPAR. Thanks are also extended to Gunnar Hornig, Antonia Wilmot-Smith, Eric Priest, Clare Parnell and Klaus Galsgaard for ongoing fruitful discussions and collaborations, as well as comments that helped to improve this manuscript.


\bibliographystyle{model2-names}

\begin{thebibliography}{97}
\expandafter\ifx\csname natexlab\endcsname\relax\def\natexlab#1{#1}\fi
\expandafter\ifx\csname url\endcsname\relax
  \def\url#1{\texttt{#1}}\fi
\expandafter\ifx\csname urlprefix\endcsname\relax\def\urlprefix{URL }\fi
\providecommand{\eprint}[2][]{\url{#2}}
\providecommand{\bibinfo}[2]{#2}
\ifx\xfnm\relax \def\xfnm[#1]{\unskip,\space#1}\fi
\bibitem[{{Al-Hachami} and {Pontin}(2010)}]{alhachami2010}
\bibinfo{author}{{Al-Hachami}, A.K.}, \bibinfo{author}{{Pontin}, D.I.},
  \bibinfo{year}{2010}.
\newblock \bibinfo{title}{{Magnetic reconnection at 3D null points: Effect of
  magnetic field asymmetry}}.
\newblock \bibinfo{journal}{Astron.~Astrophys.} \bibinfo{volume}{512},
  \bibinfo{pages}{A84}.
\bibitem[{Al-Salti and Hornig(2009)}]{alsalti2009}
\bibinfo{author}{Al-Salti, N.}, \bibinfo{author}{Hornig, G.},
  \bibinfo{year}{2009}.
\newblock \bibinfo{title}{On the solutions of three-dimensional non-null
  magnetic reconnection}.
\newblock \bibinfo{journal}{Phys. Plasmas} \bibinfo{volume}{16},
  \bibinfo{pages}{082101}.
\bibitem[{Antiochos et~al.(1999)Antiochos, DeVore and Klimchuk}]{antiochos1999}
\bibinfo{author}{Antiochos, S.K.}, \bibinfo{author}{DeVore, C.R.},
  \bibinfo{author}{Klimchuk, J.A.}, \bibinfo{year}{1999}.
\newblock \bibinfo{title}{A model for solar coronal mass ejections}.
\newblock \bibinfo{journal}{Astrophys. J.} \bibinfo{volume}{510},
  \bibinfo{pages}{485--493}.
\bibitem[{{Aulanier} et~al.(2007){Aulanier}, {Golub}, {DeLuca}, {Cirtain},
  {Kano}, {Lundquist}, {Narukage}, {Sakao} and {Weber}}]{aulanier2007}
\bibinfo{author}{{Aulanier}, G.}, \bibinfo{author}{{Golub}, L.},
  \bibinfo{author}{{DeLuca}, E.E.}, et al.,
\bibinfo{year}{2007}.
\newblock \bibinfo{title}{{Slipping magnetic reconnection in coronal loops}}.
\newblock \bibinfo{journal}{Science} \bibinfo{volume}{318},
  \bibinfo{pages}{1588}.
\bibitem[{Aulanier et~al.(2006)Aulanier, Pariat, D{\'e}moulin and
  Devore}]{aulanier2006}
\bibinfo{author}{Aulanier, G.}, \bibinfo{author}{Pariat, E.},
  \bibinfo{author}{D{\'e}moulin, P.}, \bibinfo{author}{Devore, C.R.},
  \bibinfo{year}{2006}.
\newblock \bibinfo{title}{Slip-running reconnection in quasi-separatrix
  layers}.
\newblock \bibinfo{journal}{Solar~Phys.} \bibinfo{volume}{238},
  \bibinfo{pages}{347--376}.
\bibitem[{Barnes(2007)}]{barnes2007}
\bibinfo{author}{Barnes, G.}, \bibinfo{year}{2007}.
\newblock \bibinfo{title}{On the relationship between coronal magnetic null
  points and solar eruptive events}.
\newblock \bibinfo{journal}{Astrophys. J. Lett.} \bibinfo{volume}{670},
  \bibinfo{pages}{L53--L56}.
\bibitem[{Biskamp(2000)}]{biskamp2000}
\bibinfo{author}{Biskamp, D.}, \bibinfo{year}{2000}.
\newblock \bibinfo{title}{Magnetic reconnection in plasmas}.
\newblock \bibinfo{publisher}{Cambridge University Press}.
\bibitem[{{Browning} et~al.(2008){Browning}, {Gerrard}, {Hood}, {Kevis} and
  {van der Linden}}]{browning2008}
\bibinfo{author}{{Browning}, P.K.}, \bibinfo{author}{{Gerrard}, C.},
  \bibinfo{author}{{Hood}, A.W.}, \bibinfo{author}{{Kevis}, R.},
  \bibinfo{author}{{van der Linden}, R.A.M.}, \bibinfo{year}{2008}.
\newblock \bibinfo{title}{{Heating the corona by nanoflares: simulations of
  energy release triggered by a kink instability}}.
\newblock \bibinfo{journal}{Astron.~Astrophys.} \bibinfo{volume}{485},
  \bibinfo{pages}{837--848}.
\bibitem[{Bulanov and Sakai(1997)}]{bulanov1997}
\bibinfo{author}{Bulanov, S.V.}, \bibinfo{author}{Sakai, J.},
  \bibinfo{year}{1997}.
\newblock \bibinfo{title}{Magnetic collapse in incompressible plasma flows}.
\newblock \bibinfo{journal}{J. Phys. Soc. Jpn.} \bibinfo{volume}{66},
  \bibinfo{pages}{3477--3483}.
\bibitem[{Craig and Fabling(1996)}]{craig1996}
\bibinfo{author}{Craig, I.J.D.}, \bibinfo{author}{Fabling, R.B.},
  \bibinfo{year}{1996}.
\newblock \bibinfo{title}{Exact solutions for steady-state, spine, and fan
  magnetic reconnection}.
\newblock \bibinfo{journal}{Astrophys. J.} \bibinfo{volume}{462},
  \bibinfo{pages}{969--976}.
\bibitem[{Craig and Fabling(1998)}]{craig1998}
\bibinfo{author}{Craig, I.J.D.}, \bibinfo{author}{Fabling, R.B.},
  \bibinfo{year}{1998}.
\newblock \bibinfo{title}{Dynamic magnetic reconnection in three space
  dimensions: Fan current solutions}.
\newblock \bibinfo{journal}{Phys. Plasmas} \bibinfo{volume}{5},
  \bibinfo{pages}{635--644}.
\bibitem[{Craig et~al.(1999)Craig, Fabling, Heerikhuisen and
  Watson}]{craig1999}
\bibinfo{author}{Craig, I.J.D.}, \bibinfo{author}{Fabling, R.B.},
  \bibinfo{author}{Heerikhuisen, J.}, \bibinfo{author}{Watson, P.G.},
  \bibinfo{year}{1999}.
\newblock \bibinfo{title}{Magnetic reconnection solutions in the presence of
  multiple nulls}.
\newblock \bibinfo{journal}{Astrophys. J.} \bibinfo{volume}{523},
  \bibinfo{pages}{838--848}.
\bibitem[{Craig et~al.(1995)Craig, Fabling, Henton and Rickard}]{craigetal1995}
\bibinfo{author}{Craig, I.J.D.}, \bibinfo{author}{Fabling, R.B.},
  \bibinfo{author}{Henton, S.M.}, \bibinfo{author}{Rickard, G.J.},
  \bibinfo{year}{1995}.
\newblock \bibinfo{title}{An exact solution for steady state magnetic
  reconnection in three dimensions}.
\newblock \bibinfo{journal}{Astrophys. J. Lett.} \bibinfo{volume}{455},
  \bibinfo{pages}{L197--L199}.
\bibitem[{D\a'emoulin(2006)}]{demoulin2006}
\bibinfo{author}{D\a'emoulin, P.}, \bibinfo{year}{2006}.
\newblock \bibinfo{title}{Extending the concept of separatrices to qsls for
  magnetic reconnection}.
\newblock \bibinfo{journal}{Adv.~Space Res.} \bibinfo{volume}{{37}},
  \bibinfo{pages}{1269--1282}.
\bibitem[{D\a'emoulin et~al.(1994)D\a'emoulin, H\a'enoux and
  Mandrini}]{demoulin1994}
\bibinfo{author}{D\a'emoulin, P.}, \bibinfo{author}{H\a'enoux, J.C.},
  \bibinfo{author}{Mandrini, C.H.}, \bibinfo{year}{1994}.
\newblock \bibinfo{title}{Are null magnetic points important in solar flares?}
\newblock \bibinfo{journal}{Astron. Astrophys.} \bibinfo{volume}{285},
  \bibinfo{pages}{1023}.
\bibitem[{D{\'e}moulin et~al.(1997)D{\'e}moulin, Bagala, Mandrini, H{\'e}noux
  and Rovira}]{demoulin1997}
\bibinfo{author}{D{\'e}moulin, P.}, \bibinfo{author}{Bagala, L.G.},
  \bibinfo{author}{Mandrini, C.H.}, \bibinfo{author}{H{\'e}noux, J.C.},
  \bibinfo{author}{Rovira, M.G.}, \bibinfo{year}{1997}.
\newblock \bibinfo{title}{Quasi-separatrix layers in solar flares. {II.
  Observed} magnetic configurations}.
\newblock \bibinfo{journal}{Astron.\ Astrophys.} \bibinfo{volume}{325},
  \bibinfo{pages}{305--317}.
\bibitem[{{Deng} et~al.(2009){Deng}, {Zhou}, {Li}, {Baumjohann}, {Andre},
  {Cornilleau}, {Santol{\'{\i}}k}, {Pontin}, {Reme}, {Lucek}, {Fazakerley},
  {Decreau}, {Daly}, {Nakamura}, {Tang}, {Hu}, {Pang}, {B{\"u}chner}, {Zhao},
  {Vaivads}, {Pickett}, {Ng}, {Lin}, {Fu}, {Yuan}, {Su} and {Wang}}]{deng2009}
\bibinfo{author}{{Deng}, X.H.}, \bibinfo{author}{{Zhou}, M.},
  \bibinfo{author}{{Li}, S.Y.}, et al.,
  \bibinfo{year}{2009}.
\newblock \bibinfo{title}{Dynamics and waves near multiple magnetic null points
  in reconnection diffusion region}.
\newblock \bibinfo{journal}{J.~Geophys.~Res.} \bibinfo{volume}{114},
  \bibinfo{pages}{A07216}.
\bibitem[{Dorelli et~al.(2007)Dorelli, Bhattacharjee and Raeder}]{dorelli2007}
\bibinfo{author}{Dorelli, J.C.}, \bibinfo{author}{Bhattacharjee, A.},
  \bibinfo{author}{Raeder, J.}, \bibinfo{year}{2007}.
\newblock \bibinfo{title}{Separator reconnection at earth's dayside
  magnetopause under generic northward interplanetary magnetic field
  conditions}.
\newblock \bibinfo{journal}{J.~Geophys.~Res.} \bibinfo{volume}{112},
  \bibinfo{pages}{A02202}.
\bibitem[{{Fletcher} et~al.(2001){Fletcher}, {Metcalf}, {Alexander}, {Brown}
  and {Ryder}}]{fletcher2001}
\bibinfo{author}{{Fletcher}, L.}, \bibinfo{author}{{Metcalf}, T.R.},
  \bibinfo{author}{{Alexander}, D.}, \bibinfo{author}{{Brown}, D.S.},
  \bibinfo{author}{{Ryder}, L.A.}, \bibinfo{year}{2001}.
\newblock \bibinfo{title}{{Evidence for the flare trigger site and
  three-dimensional reconnection in multiwavelength observations of a solar
  flare}}.
\newblock \bibinfo{journal}{Astrophys. J.} \bibinfo{volume}{554},
  \bibinfo{pages}{451--463}.
\bibitem[{Fukao et~al.(1975)Fukao, Ugai and Tsuda}]{fukao1975}
\bibinfo{author}{Fukao, S.}, \bibinfo{author}{Ugai, M.},
  \bibinfo{author}{Tsuda, T.}, \bibinfo{year}{1975}.
\newblock \bibinfo{title}{Topological study of magnetic field near a neutral
  point}.
\newblock \bibinfo{journal}{Rep. Ion. Sp. Res. Japan} \bibinfo{volume}{29},
  \bibinfo{pages}{133--139}.
\bibitem[{Galsgaard and Nordlund(1996)}]{galsgaard1996}
\bibinfo{author}{Galsgaard, K.}, \bibinfo{author}{Nordlund, A.},
  \bibinfo{year}{1996}.
\newblock \bibinfo{title}{Heating and activity of the solar corona: 1. Boundary
  shearing of an initially homogeneous magnetic field}.
\newblock \bibinfo{journal}{J. Geophys. Res.} \bibinfo{volume}{101},
  \bibinfo{pages}{13445--13460}.
\bibitem[{Galsgaard and Nordlund(1997)}]{galsgaard1997}
\bibinfo{author}{Galsgaard, K.}, \bibinfo{author}{Nordlund, A.},
  \bibinfo{year}{1997}.
\newblock \bibinfo{title}{Heating and activity of the solar corona: 3. Dynamics
  of a low beta plasma with 3D null points}.
\newblock \bibinfo{journal}{J. Geophys. Res.} \bibinfo{volume}{102},
  \bibinfo{pages}{231--248}.
\bibitem[{Galsgaard et~al.(2000)Galsgaard, Priest and
  Nordlund}]{galsgaardpriest2000}
\bibinfo{author}{Galsgaard, K.}, \bibinfo{author}{Priest, E.R.},
  \bibinfo{author}{Nordlund, A.}, \bibinfo{year}{2000}.
\newblock \bibinfo{title}{Three-dimensional separator reconnection - how does it
  occur?}
\newblock \bibinfo{journal}{Solar Phys.} \bibinfo{volume}{193},
  \bibinfo{pages}{1--16}.
\bibitem[{{Galsgaard} et~al.(1997){Galsgaard}, {Reddy} and
  {Rickard}}]{galsgaardreddy1997}
\bibinfo{author}{{Galsgaard}, K.}, \bibinfo{author}{{Reddy}, R.V.},
  \bibinfo{author}{{Rickard}, G.J.}, \bibinfo{year}{1997}.
\newblock \bibinfo{title}{{Energy release sites in magnetic fields containing
  single or multiple nulls}}.
\newblock \bibinfo{journal}{Solar Phys.} \bibinfo{volume}{176},
  \bibinfo{pages}{299--325}.
\bibitem[{Galsgaard et~al.(2003)Galsgaard, Titov and Neukirch}]{galsgaard2003}
\bibinfo{author}{Galsgaard, K.}, \bibinfo{author}{Titov, V.S.},
  \bibinfo{author}{Neukirch, T.}, \bibinfo{year}{2003}.
\newblock \bibinfo{title}{Magnetic pinching of hyperbolic flux tubes. ii.
  dynamic numerical model}.
\newblock \bibinfo{journal}{Astrophys.~J.} \bibinfo{volume}{595},
  \bibinfo{pages}{506--516}.
\bibitem[{{Haynes} and {Parnell}(2007)}]{haynes2007b}
\bibinfo{author}{{Haynes}, A.L.}, \bibinfo{author}{{Parnell}, C.E.},
  \bibinfo{year}{2007}.
\newblock \bibinfo{title}{{A trilinear method for finding null points in a
  three-dimensional vector space}}.
\newblock \bibinfo{journal}{Physics of Plasmas} \bibinfo{volume}{14},
  \bibinfo{pages}{082107}.
\bibitem[{{Haynes} and {Parnell}(2010)}]{haynes2010}
\bibinfo{author}{{Haynes}, A.L.}, \bibinfo{author}{{Parnell}, C.E.},
  \bibinfo{year}{2010}.
\newblock \bibinfo{title}{{A method for finding three-dimensional magnetic
  skeletons}}.
\newblock \bibinfo{journal}{Physics of Plasmas} \bibinfo{volume}{17},
  \bibinfo{pages}{092903}.
\bibitem[{{Haynes} et~al.(2007){Haynes}, {Parnell}, {Galsgaard} and
  {Priest}}]{haynes2007}
\bibinfo{author}{{Haynes}, A.L.}, \bibinfo{author}{{Parnell}, C.E.},
  \bibinfo{author}{{Galsgaard}, K.}, \bibinfo{author}{{Priest}, E.R.},
  \bibinfo{year}{2007}.
\newblock \bibinfo{title}{{Magnetohydrodynamic evolution of magnetic
  skeletons}}.
\newblock \bibinfo{journal}{Royal Society of London Proceedings Series A}
  \bibinfo{volume}{463}, \bibinfo{pages}{1097--1115}.
\bibitem[{Hendrix and {van Hoven}(1996)}]{hendrix1996}
\bibinfo{author}{Hendrix, D.}, \bibinfo{author}{{van Hoven}, G.},
  \bibinfo{year}{1996}.
\newblock \bibinfo{title}{Magnetohydrodynamic turbulence and implications for
  solar coronal heating}.
\newblock \bibinfo{journal}{Astrophys. J.} \bibinfo{volume}{467},
  \bibinfo{pages}{887--893}.
\bibitem[{Hesse(1991)}]{hesse1991}
\bibinfo{author}{Hesse, M.}, \bibinfo{year}{1991}.
\newblock \bibinfo{title}{Advances in solar system magnetohydrodynamics}.
  \bibinfo{publisher}{Cambridge University Press: Cambridge}.
\newblock p. \bibinfo{pages}{221}.
\bibitem[{Hesse and Schindler(1988)}]{hesse1988}
\bibinfo{author}{Hesse, M.}, \bibinfo{author}{Schindler, K.},
  \bibinfo{year}{1988}.
\newblock \bibinfo{title}{A theoretical foundation of general magnetic
  reconnection}.
\newblock \bibinfo{journal}{J. Geophys. Res.} \bibinfo{volume}{93},
  \bibinfo{pages}{5558--5567}.
\bibitem[{{Hood} et~al.(2009){Hood}, {Browning} and {van der
  Linden}}]{hood2009}
\bibinfo{author}{{Hood}, A.W.}, \bibinfo{author}{{Browning}, P.K.},
  \bibinfo{author}{{van der Linden}, R.A.M.}, \bibinfo{year}{2009}.
\newblock \bibinfo{title}{{Coronal heating by magnetic reconnection in loops
  with zero net current}}.
\newblock \bibinfo{journal}{Astron.~Astrophys.} \bibinfo{volume}{506},
  \bibinfo{pages}{913--925}.
\bibitem[{Hornig(2001)}]{hornig2001}
\bibinfo{author}{Hornig, G.}, \bibinfo{year}{2001}.
\newblock \bibinfo{title}{The geometry of reconnection, in: Ricca, R.L. (Ed.), An introduction to the geometry and topology of fluid  flows}. \bibinfo{publisher}{Kluwer, Dordrecht}.
\newblock pp. \bibinfo{pages}{295--313}.
\bibitem[{Hornig(2007a)}]{hornig2007a}
\bibinfo{author}{Hornig, G.}, \bibinfo{year}{2007}a.
\newblock \bibinfo{title}{Fundamental concepts, in: Birn, J. and Priest, E.R. (Eds.), Reconnection of magnetic fields}.
  \bibinfo{publisher}{Cambridge University Press}. 
\newblock pp. \bibinfo{pages}{25--45}.
\bibitem[{Hornig(2007b)}]{hornig2007b}
\bibinfo{author}{Hornig, G.}, \bibinfo{year}{2007}b.
\newblock \bibinfo{title}{3D reconnection without nulls, in: Birn, J. and Priest, E.R. (Eds.), Reconnection of magnetic fields}.
  \bibinfo{publisher}{Cambridge University Press}. 
  \newblock pp. \bibinfo{pages}{45--62}.
\bibitem[{Hornig and Priest(2003)}]{hornig2003}
\bibinfo{author}{Hornig, G.}, \bibinfo{author}{Priest, E.R.},
  \bibinfo{year}{2003}.
\newblock \bibinfo{title}{Evolution of magnetic flux in an isolated
  reconnection process}.
\newblock \bibinfo{journal}{Phys. Plasmas} \bibinfo{volume}{10},
  \bibinfo{pages}{2712--1721}.
\bibitem[{Hornig and Schindler(1996)}]{hornig1996}
\bibinfo{author}{Hornig, G.}, \bibinfo{author}{Schindler, K.},
  \bibinfo{year}{1996}.
\newblock \bibinfo{title}{Magnetic topology and the problem of its invariant
  definition}.
\newblock \bibinfo{journal}{Phys. Plasmas} \bibinfo{volume}{3},
  \bibinfo{pages}{781--791}.
\bibitem[{Klapper et~al.(1996)Klapper, Rado and Tabor}]{klapper1996}
\bibinfo{author}{Klapper, I.}, \bibinfo{author}{Rado, A.},
  \bibinfo{author}{Tabor, M.}, \bibinfo{year}{1996}.
\newblock \bibinfo{title}{A Lagrangian study of dynamics and singularity
  formation at magnetic null points in ideal three-dimensional
  magnetohydrodynamics}.
\newblock \bibinfo{journal}{Phys. Plasmas} \bibinfo{volume}{3},
  \bibinfo{pages}{4281--4283}.
\bibitem[{{Kliem} et~al.(2004){Kliem}, {Titov} and {T{\"o}r{\"o}k}}]{kliem2004}
\bibinfo{author}{{Kliem}, B.}, \bibinfo{author}{{Titov}, V.S.},
  \bibinfo{author}{{T{\"o}r{\"o}k}, T.}, \bibinfo{year}{2004}.
\newblock \bibinfo{title}{{Formation of current sheets and sigmoidal structure
  by the kink instability of a magnetic loop}}.
\newblock \bibinfo{journal}{Astron.\ Astrophys.} \bibinfo{volume}{413},
  \bibinfo{pages}{L23--L26}.
\bibitem[{Lau and Finn(1990)}]{lau1990}
\bibinfo{author}{Lau, Y.T.}, \bibinfo{author}{Finn, J.M.},
  \bibinfo{year}{1990}.
\newblock \bibinfo{title}{Three dimensional kinematic reconnection in the
  presence of field nulls and closed field lines}.
\newblock \bibinfo{journal}{Astrophys. J.} \bibinfo{volume}{350},
  \bibinfo{pages}{672--691}.
\bibitem[{Linton et~al.(2001)Linton, Dahlburg and Antiochos}]{linton2001}
\bibinfo{author}{Linton, M.}, \bibinfo{author}{Dahlburg, R.B.},
  \bibinfo{author}{Antiochos, S.K.}, \bibinfo{year}{2001}.
\newblock \bibinfo{title}{Reconnection of twisted flux tubes as a function of
  contact angle}.
\newblock \bibinfo{journal}{Astrophys. J.} \bibinfo{volume}{553},
  \bibinfo{pages}{905--921}.
\bibitem[{Linton and Priest(2003)}]{linton2003}
\bibinfo{author}{Linton, M.}, \bibinfo{author}{Priest, E.R.},
  \bibinfo{year}{2003}.
\newblock \bibinfo{title}{Three-dimensional reconnection of untwisted flux
  tubes}.
\newblock \bibinfo{journal}{Astrophys. J.} \bibinfo{volume}{595},
  \bibinfo{pages}{1259--1276}.
\bibitem[{Longcope(1996)}]{longcope1996}
\bibinfo{author}{Longcope, D.W.}, \bibinfo{year}{1996}.
\newblock \bibinfo{title}{Topology and current ribbons: A model for current,
  reconnection and flaring in a complex, evolving corona}.
\newblock \bibinfo{journal}{Solar Phys.} \bibinfo{volume}{169},
  \bibinfo{pages}{91--121}.
\bibitem[{Longcope and Cowley(1996)}]{longcopecowley1996}
\bibinfo{author}{Longcope, D.W.}, \bibinfo{author}{Cowley, S.C.},
  \bibinfo{year}{1996}.
\newblock \bibinfo{title}{Current sheet formation along three-dimensional
  magnetic separators}.
\newblock \bibinfo{journal}{Phys. Plasmas} \bibinfo{volume}{3},
  \bibinfo{pages}{2885--2897}.
\bibitem[{Longcope et~al.(2005)Longcope, McKenzie, Cirtain and
  Scott}]{longcope2005a}
\bibinfo{author}{Longcope, D.W.}, \bibinfo{author}{McKenzie, D.},
  \bibinfo{author}{Cirtain, J.}, \bibinfo{author}{Scott, J.},
  \bibinfo{year}{2005}.
\newblock \bibinfo{title}{Observations of separator reconnection to an emerging
  active region}.
\newblock \bibinfo{journal}{Astrophys. J.} \bibinfo{volume}{630},
  \bibinfo{pages}{569}.
\bibitem[{{Longcope} and {Parnell}(2009)}]{longcope2009}
\bibinfo{author}{{Longcope}, D.W.}, \bibinfo{author}{{Parnell}, C.E.},
  \bibinfo{year}{2009}.
\newblock \bibinfo{title}{{The number of magnetic null points in the quiet sun corona}}.
\newblock \bibinfo{journal}{Solar Phys.} \bibinfo{volume}{254},
  \bibinfo{pages}{51--75}.
\bibitem[{{Longcope} and {Strauss}(1994)}]{longcope1994}
\bibinfo{author}{{Longcope}, D.W.}, \bibinfo{author}{{Strauss}, H.R.},
  \bibinfo{year}{1994}.
\newblock \bibinfo{title}{{The form of ideal current layers in line-tied
  magnetic fields}}.
\newblock \bibinfo{journal}{Astrophys.~J.} \bibinfo{volume}{437},
  \bibinfo{pages}{851--859}.
\bibitem[{Luoni et~al.(2007)Luoni, Mandrini, Cristiani and D{\'
  e}moulin}]{luoni2007}
\bibinfo{author}{Luoni, M.L.}, \bibinfo{author}{Mandrini, H.H.},
  \bibinfo{author}{Cristiani, G.D.}, \bibinfo{author}{D{\' e}moulin, P.},
  \bibinfo{year}{2007}.
\newblock \bibinfo{title}{The magnetic field topology associated with two M-flares}.
\newblock \bibinfo{journal}{Adv. Space Res.} \bibinfo{volume}{39},
  \bibinfo{pages}{1382--1388}.
\bibitem[{Lynch et~al.(2008)Lynch, Antiochos, DeVore, Luhmann and
  Zurbuchen}]{Lynch2008}
\bibinfo{author}{Lynch, G.J.}, \bibinfo{author}{Antiochos, S.K.},
  \bibinfo{author}{DeVore, C.R.}, \bibinfo{author}{Luhmann, J.G.},
  \bibinfo{author}{Zurbuchen, T.H.}, \bibinfo{year}{2008}.
\newblock \bibinfo{title}{Topological evolution of a fast magnetic breakout CME
  in three dimensions}.
\newblock \bibinfo{journal}{Astrophys.~J.} \bibinfo{volume}{683},
  \bibinfo{pages}{1192--1206}.
\bibitem[{Mandrini et~al.(2006)Mandrini, D{\' e}moulin, Schmieder, Deluca,
  Pariat and Uddin}]{mandrini2006}
\bibinfo{author}{Mandrini, C.H.}, \bibinfo{author}{D{\' e}moulin, P.},
  \bibinfo{author}{Schmieder, B.}, \bibinfo{author}{Deluca, E.E.},
  \bibinfo{author}{Pariat, E.}, \bibinfo{author}{Uddin, W.},
  \bibinfo{year}{2006}.
\newblock \bibinfo{title}{Companion event and precursor of the X17 flare on 28
  October 2003}.
\newblock \bibinfo{journal}{Solar Phys.} \bibinfo{volume}{238},
  \bibinfo{pages}{293--312}.
\bibitem[{Masson et~al.(2009)Masson, Pariat, Aulanier and
  Schrijver}]{masson2009}
\bibinfo{author}{Masson, S.}, \bibinfo{author}{Pariat, E.},
  \bibinfo{author}{Aulanier, G.}, \bibinfo{author}{Schrijver, C.J.},
  \bibinfo{year}{2009}.
\newblock \bibinfo{title}{The nature of flare ribbons in coronal null-point
  topology}.
\newblock \bibinfo{journal}{Astrophys.\ J.} \bibinfo{volume}{700},
  \bibinfo{pages}{559--578}.
\bibitem[{McLaughlin et~al.(2010)McLaughlin, Hood and
  de~Moortel}]{mclaughlin2010}
\bibinfo{author}{McLaughlin, J.A.}, \bibinfo{author}{Hood, A.W.},
  \bibinfo{author}{de~Moortel, I.}, \bibinfo{year}{2010}.
\newblock \bibinfo{title}{Review article: MHD wave propagation near coronal
  null points of magnetic fields}.
\newblock \bibinfo{journal}{Space~Sci.~Rev.} , \bibinfo{pages}{62}.
\bibitem[{Mellor et~al.(2003)Mellor, Titov and Priest}]{mellor2003}
\bibinfo{author}{Mellor, C.}, \bibinfo{author}{Titov, V.S.},
  \bibinfo{author}{Priest, E.R.}, \bibinfo{year}{2003}.
\newblock \bibinfo{title}{Linear collapse of spatially linear 3D potential null
  points.}
\newblock \bibinfo{journal}{Geophys. Astrophys. Fluid Dynamics}
  \bibinfo{volume}{97}, \bibinfo{pages}{489--505}.
\bibitem[{{Ng} and {Bhattacharjee}(1998)}]{ng1998}
\bibinfo{author}{{Ng}, C.S.}, \bibinfo{author}{{Bhattacharjee}, A.},
  \bibinfo{year}{1998}.
\newblock \bibinfo{title}{{Nonequilibrium and current sheet formation in
  line-tied magnetic fields}}.
\newblock \bibinfo{journal}{Phys. Plasmas} \bibinfo{volume}{5},
  \bibinfo{pages}{4028--4040}.
\bibitem[{{Pariat} et~al.(2009){Pariat}, {Antiochos} and {DeVore}}]{pariat2009}
\bibinfo{author}{{Pariat}, E.}, \bibinfo{author}{{Antiochos}, S.K.},
  \bibinfo{author}{{DeVore}, C.R.}, \bibinfo{year}{2009}.
\newblock \bibinfo{title}{{A model for solar polar jets}}.
\newblock \bibinfo{journal}{Astrophys. J.} \bibinfo{volume}{691},
  \bibinfo{pages}{61--74}.
\bibitem[{Parker(1972)}]{parker1972}
\bibinfo{author}{Parker, E.N.}, \bibinfo{year}{1972}.
\newblock \bibinfo{title}{Topological dissipation and the small-scale fields in
  turbulent gases}.
\newblock \bibinfo{journal}{Astrophys.~J.} \bibinfo{volume}{174},
  \bibinfo{pages}{499}.
\bibitem[{Parnell et~al.(2008)Parnell, Haynes and Galsgaard}]{parnell2008}
\bibinfo{author}{Parnell, C.E.}, \bibinfo{author}{Haynes, A.L.},
  \bibinfo{author}{Galsgaard, K.}, \bibinfo{year}{2008}.
\newblock \bibinfo{title}{Recursive reconnection and magnetic skeletons}.
\newblock \bibinfo{journal}{Astrophys.~J.} \bibinfo{volume}{675},
  \bibinfo{pages}{1656--1667}.
\bibitem[{Parnell et~al.(2010)Parnell, Haynes and Galsgaard}]{parnell2010}
\bibinfo{author}{Parnell, C.E.}, \bibinfo{author}{Haynes, A.L.},
  \bibinfo{author}{Galsgaard, K.}, \bibinfo{year}{2010}.
\newblock \bibinfo{title}{Structure of magnetic separators and separator
  reconnection}.
\newblock \bibinfo{journal}{J.~Geophys.~Res.} \bibinfo{volume}{115},
  \bibinfo{pages}{A02102}.
\bibitem[{Parnell et~al.(1996)Parnell, Smith, Neukirch and
  Priest}]{parnell1996}
\bibinfo{author}{Parnell, C.E.}, \bibinfo{author}{Smith, J.M.},
  \bibinfo{author}{Neukirch, T.}, \bibinfo{author}{Priest, E.R.},
  \bibinfo{year}{1996}.
\newblock \bibinfo{title}{The structure of three-dimensional magnetic neutral
  points}.
\newblock \bibinfo{journal}{Phys.~Plasmas} \bibinfo{volume}{3},
  \bibinfo{pages}{759--770}.
\bibitem[{Pontin et~al.(2007a)Pontin, Bhattacharjee and
  Galsgaard}]{pontinbhat2007a}
\bibinfo{author}{Pontin, D.I.}, \bibinfo{author}{Bhattacharjee, A.},
  \bibinfo{author}{Galsgaard, K.}, \bibinfo{year}{2007}a.
\newblock \bibinfo{title}{Current sheet formation and non-ideal behaviour at
  three-dimensional magnetic null points}.
\newblock \bibinfo{journal}{Phys.~Plasmas} \bibinfo{volume}{14},
  \bibinfo{pages}{052106}.
\bibitem[{Pontin et~al.(2007b)Pontin, Bhattacharjee and
  Galsgaard}]{pontinbhat2007b}
\bibinfo{author}{Pontin, D.I.}, \bibinfo{author}{Bhattacharjee, A.},
  \bibinfo{author}{Galsgaard, K.}, \bibinfo{year}{2007}b.
\newblock \bibinfo{title}{Current sheets at three-dimensional magnetic null
  points: Effect of compressibility}.
\newblock \bibinfo{journal}{Phys.~Plasmas} \bibinfo{volume}{14},
  \bibinfo{pages}{052109}.
\bibitem[{Pontin and Craig(2005)}]{pontincraig2005}
\bibinfo{author}{Pontin, D.I.}, \bibinfo{author}{Craig, I.J.D.},
  \bibinfo{year}{2005}.
\newblock \bibinfo{title}{Current singularities at finitely compressible
  three-dimensional magnetic null points}.
\newblock \bibinfo{journal}{Phys.~Plasmas} \bibinfo{volume}{12},
  \bibinfo{pages}{072112}.
\bibitem[{Pontin and Craig(2006)}]{pontin2006}
\bibinfo{author}{Pontin, D.I.}, \bibinfo{author}{Craig, I.J.D.},
  \bibinfo{year}{2006}.
\newblock \bibinfo{title}{Dynamic 3D reconnection in a separator geometry with
  two null points}.
\newblock \bibinfo{journal}{Astrophys.~J.} \bibinfo{volume}{642},
  \bibinfo{pages}{568--578}.
\bibitem[{Pontin and Galsgaard(2007)}]{pontingalsgaard2007}
\bibinfo{author}{Pontin, D.I.}, \bibinfo{author}{Galsgaard, K.},
  \bibinfo{year}{2007}.
\newblock \bibinfo{title}{Current amplification and magnetic reconnection at a
  3D null point. Physical characteristics}.
\newblock \bibinfo{journal}{J.~Geophys.~Res.} \bibinfo{volume}{112},
  \bibinfo{pages}{A03103}.
\bibitem[{Pontin et~al.(2005a)Pontin, Galsgaard, Hornig and
  Priest}]{pontingalsgaard2005}
\bibinfo{author}{Pontin, D.I.}, \bibinfo{author}{Galsgaard, K.},
  \bibinfo{author}{Hornig, G.}, \bibinfo{author}{Priest, E.R.},
  \bibinfo{year}{2005}a.
\newblock \bibinfo{title}{A fully magnetohydrodynamic simulation of 3D non-null
  reconnection}.
\newblock \bibinfo{journal}{Phys.~Plasmas} \bibinfo{volume}{12},
  \bibinfo{pages}{052307}.
\bibitem[{Pontin et~al.(2004)Pontin, Hornig and Priest}]{pontin2004}
\bibinfo{author}{Pontin, D.I.}, \bibinfo{author}{Hornig, G.},
  \bibinfo{author}{Priest, E.R.}, \bibinfo{year}{2004}.
\newblock \bibinfo{title}{Kinematic reconnection at a magnetic null point:
  Spine-aligned current}.
\newblock \bibinfo{journal}{Geophys. Astrophys. Fluid Dynamics}
  \bibinfo{volume}{98}, \bibinfo{pages}{407--428}.
\bibitem[{Pontin et~al.(2005b)Pontin, Hornig and Priest}]{pontinhornig2005}
\bibinfo{author}{Pontin, D.I.}, \bibinfo{author}{Hornig, G.},
  \bibinfo{author}{Priest, E.R.}, \bibinfo{year}{2005}b.
\newblock \bibinfo{title}{Kinematic reconnection at a magnetic null point:
  Fan-aligned current}.
\newblock \bibinfo{journal}{Geophys. Astrophys. Fluid Dynamics}
  \bibinfo{volume}{99}, \bibinfo{pages}{77--93}.
\bibitem[{Pontin et~al.(2011)Pontin, Wilmot-Smith, Hornig and
  Galsgaard}]{pontin2010}
\bibinfo{author}{Pontin, D.I.}, \bibinfo{author}{Wilmot-Smith, A.L.},
  \bibinfo{author}{Hornig, G.}, \bibinfo{author}{Galsgaard, K.},
  \bibinfo{year}{2011}.
\newblock \bibinfo{title}{Dynamics of braided coronal loops. II.~Cascade to
  multiple small-scale reconnection events}.
\newblock \bibinfo{journal}{Astron. Astrophys.}
  \bibinfo{volume}{525}, \bibinfo{pages}{A57}.
\bibitem[{Priest and D\a'emoulin(1995)}]{priest1995}
\bibinfo{author}{Priest, E.R.}, \bibinfo{author}{D\a'emoulin, P.},
  \bibinfo{year}{1995}.
\newblock \bibinfo{title}{Three-dimensional magnetic reconnection without null
  points. 1. Basic theory of magnetic flipping}.
\newblock \bibinfo{journal}{J. Geophys. Res.} \bibinfo{volume}{100},
  \bibinfo{pages}{23443--23463}.
\bibitem[{Priest and Forbes(1992)}]{priest1992}
\bibinfo{author}{Priest, E.R.}, \bibinfo{author}{Forbes, T.G.},
  \bibinfo{year}{1992}.
\newblock \bibinfo{title}{Magnetic flipping - reconnection in three dimensions
  without null points}.
\newblock \bibinfo{journal}{J. Geophys. Res.} \bibinfo{volume}{97},
  \bibinfo{pages}{1521--1531}.
\bibitem[{Priest and Forbes(2000)}]{priest2000}
\bibinfo{author}{Priest, E.R.}, \bibinfo{author}{Forbes, T.G.},
  \bibinfo{year}{2000}.
\newblock \bibinfo{title}{Magnetic reconnection: MHD theory and applications}.
\newblock \bibinfo{publisher}{Cambridge University Press, Cambridge}.
\bibitem[{Priest et~al.(2003)Priest, Hornig and Pontin}]{priesthornig2003}
\bibinfo{author}{Priest, E.R.}, \bibinfo{author}{Hornig, G.},
  \bibinfo{author}{Pontin, D.I.}, \bibinfo{year}{2003}.
\newblock \bibinfo{title}{On the nature of three-dimensional magnetic
  reconnection}.
\newblock \bibinfo{journal}{J. Geophys. Res.} \bibinfo{volume}{108},
  \bibinfo{pages}{1285}.
\bibitem[{Priest and Pontin(2009)}]{priest2009}
\bibinfo{author}{Priest, E.R.}, \bibinfo{author}{Pontin, D.I.},
  \bibinfo{year}{2009}.
\newblock \bibinfo{title}{Three-dimensional null point reconnection regimes}.
\newblock \bibinfo{journal}{Phys.~Plasmas} \bibinfo{volume}{16},
  \bibinfo{pages}{122101}.
\bibitem[{Priest and Titov(1996)}]{priest1996}
\bibinfo{author}{Priest, E.R.}, \bibinfo{author}{Titov, V.S.},
  \bibinfo{year}{1996}.
\newblock \bibinfo{title}{Magnetic reconnection at three-dimensional null
  points}.
\newblock \bibinfo{journal}{Phil. Trans. R. Soc. A} \bibinfo{volume}{354},
  \bibinfo{pages}{2951--2992}.
\bibitem[{Rappazzo et~al.(2008)Rappazzo, Velli, Einaudi and
  Dahlburg}]{rappazzo2008}
\bibinfo{author}{Rappazzo, A.F.}, \bibinfo{author}{Velli, M.},
  \bibinfo{author}{Einaudi, G.}, \bibinfo{author}{Dahlburg, R.B.},
  \bibinfo{year}{2008}.
\newblock \bibinfo{title}{Nonlinear dynamics of the Parker scenario for coronal
  heating}.
\newblock \bibinfo{journal}{Astrophys. J.} \bibinfo{volume}{677},
  \bibinfo{pages}{1348--1366}.
\bibitem[{{R{\'e}gnier} et~al.(2008){R{\'e}gnier}, {Parnell} and
  {Haynes}}]{regnier2008}
\bibinfo{author}{{R{\'e}gnier}, S.}, \bibinfo{author}{{Parnell}, C.E.},
  \bibinfo{author}{{Haynes}, A.L.}, \bibinfo{year}{2008}.
\newblock \bibinfo{title}{{A new view of quiet-Sun topology from Hinode/SOT}}.
\newblock \bibinfo{journal}{Astron. Astrophys.} \bibinfo{volume}{484},
  \bibinfo{pages}{L47--L50}.
\bibitem[{Rickard and Titov(1996)}]{rickard1996}
\bibinfo{author}{Rickard, G.J.}, \bibinfo{author}{Titov, V.S.},
  \bibinfo{year}{1996}.
\newblock \bibinfo{title}{Current accumulation at a three-dimensional magnetic
  null}.
\newblock \bibinfo{journal}{Astrophys. J.} \bibinfo{volume}{472},
  \bibinfo{pages}{840--852}.
\bibitem[{Schindler et~al.(1988)Schindler, Hesse and Birn}]{schindler1988}
\bibinfo{author}{Schindler, K.}, \bibinfo{author}{Hesse, M.},
  \bibinfo{author}{Birn, J.}, \bibinfo{year}{1988}.
\newblock \bibinfo{title}{General magnetic reconnection, parallel electric
  fields, and helicity}.
\newblock \bibinfo{journal}{J.~Geophys.~Res.} \bibinfo{volume}{93},
  \bibinfo{pages}{5547--5557}.
\bibitem[{Titov(2007)}]{titov2007}
\bibinfo{author}{Titov, V.S.}, \bibinfo{year}{2007}.
\newblock \bibinfo{title}{Generalized squashing factors for covariant
  description of magnetic connectivity in the solar corona}.
\newblock \bibinfo{journal}{Astrophys.\ J.} \bibinfo{volume}{660},
  \bibinfo{pages}{863--873}.
\bibitem[{{Titov} and {D{\'e}moulin}(1999)}]{titov1999}
\bibinfo{author}{{Titov}, V.S.}, \bibinfo{author}{{D{\'e}moulin}, P.},
  \bibinfo{year}{1999}.
\newblock \bibinfo{title}{{Basic topology of twisted magnetic configurations in
  solar flares}}.
\newblock \bibinfo{journal}{Astron.\ Astrophys.} \bibinfo{volume}{351},
  \bibinfo{pages}{707--720}.
\bibitem[{Titov et~al.(2009)Titov, Forbes, Priest, Miki{\'c} and
  Linker}]{titov2009}
\bibinfo{author}{Titov, V.S.}, \bibinfo{author}{Forbes, T.G.},
  \bibinfo{author}{Priest, E.R.}, \bibinfo{author}{Miki{\'c}, Z.},
  \bibinfo{author}{Linker, J.A.}, \bibinfo{year}{2009}.
\newblock \bibinfo{title}{Slip-squashing factors as a measure of
  three-dimensional magnetic reconnection}.
\newblock \bibinfo{journal}{Astrophys.\ J.} \bibinfo{volume}{693},
  \bibinfo{pages}{1029--1044}.
\bibitem[{Titov et~al.(2002)Titov, Hornig and D\a'emoulin}]{titov2002}
\bibinfo{author}{Titov, V.S.}, \bibinfo{author}{Hornig, G.},
  \bibinfo{author}{D\a'emoulin, P.}, \bibinfo{year}{2002}.
\newblock \bibinfo{title}{The theory of magnetic connectivity in the corona}.
\newblock \bibinfo{journal}{J. Geophys. Res.} \bibinfo{volume}{107},
  \bibinfo{pages}{SSH 3--1}.
\bibitem[{{Titov} et~al.(2008){Titov}, {Mikic}, {Linker} and
  {Lionello}}]{titov2008}
\bibinfo{author}{{Titov}, V.S.}, \bibinfo{author}{{Mikic}, Z.},
  \bibinfo{author}{{Linker}, J.A.}, \bibinfo{author}{{Lionello}, R.},
  \bibinfo{year}{2008}.
\newblock \bibinfo{title}{{1997 May 12 coronal mass ejection event. I. A
  simplified model of the preeruptive magnetic structure}}.
\newblock \bibinfo{journal}{Astrophys.\ J.} \bibinfo{volume}{675},
  \bibinfo{pages}{1614--1628}.
\bibitem[{Titov et~al.(2004)Titov, Tassi and Hornig}]{titov2004}
\bibinfo{author}{Titov, V.S.}, \bibinfo{author}{Tassi, E.},
  \bibinfo{author}{Hornig, G.}, \bibinfo{year}{2004}.
\newblock \bibinfo{title}{Exact solutions for steady reconnective annihilation
  revisited}.
\newblock \bibinfo{journal}{Phys. Plasmas} \bibinfo{volume}{11},
  \bibinfo{pages}{4662--4671}.
\bibitem[{T\"or\"ok et~al.(2009)T\"or\"ok, Aulanier, Schmieder, Reeves and
  Golub}]{torok2009}
\bibinfo{author}{T\"or\"ok, T.}, \bibinfo{author}{Aulanier, G.},
  \bibinfo{author}{Schmieder, B.}, \bibinfo{author}{Reeves, K.K.},
  \bibinfo{author}{Golub, L.}, \bibinfo{year}{2009}.
\newblock \bibinfo{title}{Fan-spine topology formation through two-step
  reconnection driven by twisted flux emergence}.
\newblock \bibinfo{journal}{Astrophys.\ J.} \bibinfo{volume}{704},
  \bibinfo{pages}{485--495}.
\bibitem[{Ugarte-Urra et~al.(2007)Ugarte-Urra, Warren and
  Winebarger}]{UgarteUrra2007}
\bibinfo{author}{Ugarte-Urra, I.}, \bibinfo{author}{Warren, H.P.},
  \bibinfo{author}{Winebarger, A.R.}, \bibinfo{year}{2007}.
\newblock \bibinfo{title}{The magnetic topology of coronal mass ejection
  sources}.
\newblock \bibinfo{journal}{Astrophys. J.} \bibinfo{volume}{662},
  \bibinfo{pages}{1293--1301}.
\bibitem[{{van Ballegooijen}(1985)}]{vanballegooijen1985}
\bibinfo{author}{{van Ballegooijen}, A.A.}, \bibinfo{year}{1985}.
\newblock \bibinfo{title}{Electric currents in the solar corona and the
  existence of magnetostatic equilibrium}.
\newblock \bibinfo{journal}{Astrophys.~J.} \bibinfo{volume}{298},
  \bibinfo{pages}{421}.
\bibitem[{Wilmot-Smith et~al.(2009a)Wilmot-Smith, Hornig and
  Pontin}]{wilmotsmith2009a}
\bibinfo{author}{Wilmot-Smith, A.L.}, \bibinfo{author}{Hornig, G.},
  \bibinfo{author}{Pontin, D.I.}, \bibinfo{year}{2009}a.
\newblock \bibinfo{title}{Magnetic braiding and parallel electric fields}.
\newblock \bibinfo{journal}{Astrophys. J.} \bibinfo{volume}{696},
  \bibinfo{pages}{1339--1347}.
\bibitem[{Wilmot-Smith et~al.(2009b)Wilmot-Smith, Hornig and
  Pontin}]{wilmotsmith2009b}
\bibinfo{author}{Wilmot-Smith, A.L.}, \bibinfo{author}{Hornig, G.},
  \bibinfo{author}{Pontin, D.I.}, \bibinfo{year}{2009}b.
\newblock \bibinfo{title}{Magnetic braiding and quasi-separatrix layers}.
\newblock \bibinfo{journal}{Astrophys. J.} \bibinfo{volume}{704},
  \bibinfo{pages}{1288--1295}.
\bibitem[{Wilmot-Smith et~al.(2006)Wilmot-Smith, Hornig and
  Priest}]{wilmotsmith2006}
\bibinfo{author}{Wilmot-Smith, A.L.}, \bibinfo{author}{Hornig, G.},
  \bibinfo{author}{Priest, E.R.}, \bibinfo{year}{2006}.
\newblock \bibinfo{title}{Dynamic non-null magnetic reconnection in three
  dimensions. I. Particular solutions}.
\newblock \bibinfo{journal}{Proc. R. Soc. A} \bibinfo{volume}{{462}},
  \bibinfo{pages}{2877--2895}.
\bibitem[{Wilmot-Smith et~al.(2009c)Wilmot-Smith, Hornig and
  Priest}]{wilmotsmith2009c}
\bibinfo{author}{Wilmot-Smith, A.L.}, \bibinfo{author}{Hornig, G.},
  \bibinfo{author}{Priest, E.R.}, \bibinfo{year}{2009}c.
\newblock \bibinfo{title}{Dynamic non-null magnetic reconnection in three
  dimensions. II. Composite solutions}.
\newblock \bibinfo{journal}{Geophys.~Astrophys.~Fluid Dynamics}
  \bibinfo{volume}{103}, \bibinfo{pages}{515--534}.
\bibitem[{Wilmot-Smith et~al.(2010)Wilmot-Smith, Pontin and
  Hornig}]{wilmotsmith2010}
\bibinfo{author}{Wilmot-Smith, A.L.}, \bibinfo{author}{Pontin, D.I.},
  \bibinfo{author}{Hornig, G.}, \bibinfo{year}{2010}.
\newblock \bibinfo{title}{Dynamics of braided coronal loops - I.~Onset of
  magnetic reconnection}.
\newblock \bibinfo{journal}{Astron. Astrophys.} \bibinfo{volume}{516},
  \bibinfo{pages}{A5}.
\bibitem[{Wyper and Jain(2010)}]{wyper2010}
\bibinfo{author}{Wyper, P.}, \bibinfo{author}{Jain, R.}, \bibinfo{year}{2010}.
\newblock \bibinfo{title}{Torsional magnetic reconnection at three dimensional
  null points: A phenomenological study}.
\newblock \bibinfo{journal}{Phys.~Plasmas} \bibinfo{volume}{17},
  \bibinfo{pages}{092902}.
\bibitem[{{Xiao} et~al.(2007){Xiao}, {Wang}, {Pu}, {Ma}, {Zhao}, {Zhou},
  {Wang}, {Kivelson}, {Fu}, {Liu}, {Zong}, {Dunlop}, {Glassmeier}, {Lucek},
  {Reme}, {Dandouras} and {Escoubet}}]{xiao2007}
\bibinfo{author}{{Xiao}, C.J.}, \bibinfo{author}{{Wang}, X.G.},
  \bibinfo{author}{{Pu}, Z.Y.}, et al.,
  \bibinfo{year}{2007}.
\newblock \bibinfo{title}{{Satellite observations of separator-line geometry of
  three-dimensional magnetic reconnection}}.
\newblock \bibinfo{journal}{Nature Physics} \bibinfo{volume}{3},
  \bibinfo{pages}{609--613}.
\bibitem[{Xiao et~al.(2006)Xiao, Wang, Pu, Zhao, Wang, Ma, Fu, Kivelson, Liu,
  Zong, Glassmeier, Balogh, Korth, Reme and Escoubet}]{xiao2006}
\bibinfo{author}{Xiao, C.J.}, \bibinfo{author}{Wang, X.G.},
  \bibinfo{author}{Pu, Z.Y.}, et al.,
  \bibinfo{year}{2006}.
\newblock \bibinfo{title}{In-situ evidence for the structure of the magnetic
  null in a 3D reconnection event in the Earth's magnetotail}.
\newblock \bibinfo{journal}{Nature Physics} \bibinfo{volume}{2},
  \bibinfo{pages}{478--483}.
\bibitem[{{Yamada} et~al.(2010){Yamada}, {Kulsrud} and {Ji}}]{yamada2010}
\bibinfo{author}{{Yamada}, M.}, \bibinfo{author}{{Kulsrud}, R.},
  \bibinfo{author}{{Ji}, H.}, \bibinfo{year}{2010}.
\newblock \bibinfo{title}{{Magnetic reconnection}}.
\newblock \bibinfo{journal}{Reviews of Modern Physics} \bibinfo{volume}{82},
  \bibinfo{pages}{603--664}.
\bibitem[{{Zweibel} and {Yamada}(2009)}]{zweibel2009}
\bibinfo{author}{{Zweibel}, E.G.}, \bibinfo{author}{{Yamada}, M.},
  \bibinfo{year}{2009}.
\newblock \bibinfo{title}{{Magnetic reconnection in astrophysical and
  laboratory plasmas}}.
\newblock \bibinfo{journal}{Ann.\ Rev.\ Astron.\ Astrophys.}
  \bibinfo{volume}{47}, \bibinfo{pages}{291--332}.

\end{thebibliography}

%

\end{document}